\documentclass{aa}  

\usepackage{graphicx}
\usepackage{txfonts}

\usepackage{natbib} 
\usepackage[table]{xcolor}

\begin{document}

\renewcommand{\arraystretch}{1.2}

\title{The EPOCH Project: I. Periodic variable stars in the EROS-2 LMC database}

\author{Dae-Won Kim\inst{1}\and
Pavlos Protopapas\inst{2}\and
Coryn A.L. Bailer-Jones\inst{1}\and
Yong-Ik Byun\inst{3}\and
Seo-Won Chang\inst{3}\and
Jean-Baptiste Marquette\inst{4}\and
Min-Su Shin\inst{5}}
\institute{Max-Planck Institute for Astronomy, K\"{o}nigstuhl 17, D-69117 Heidelberg, Germany\and
Harvard-Smithsonian Center for Astrophysics, 60 Garden, Cambridge, MA 02138, USA\and
Department of Astronomy and University Observatory, Yonsei University, 50 Yonsei-Ro, Seoul 120-749, South Korea\and
UPMC-CNRS, UMR7095, Institut d'Astrophysique de Paris, F-75014, Paris, France\and
Department of Physics, Oxford University, Parks Road, Oxford OX1 3PU, UK
}

\abstract {
The EPOCH (EROS-2 periodic variable star classification using machine learning) project
aims to detect periodic variable stars in the  EROS-2 light curve database.
In this paper, we present the first result of the
classification of periodic variable stars in the EROS-2 LMC database.
To classify these variables, we first built a
training set by compiling known variables
in the Large Magellanic Cloud area from the OGLE and MACHO surveys.
We crossmatched these variables with the EROS-2 sources
and extracted 22 variability features 
from 28\,392 light curves of the corresponding EROS-2 sources.
We then used the random forest method to classify the
EROS-2 sources in the training set.
We designed the model to separate not only
$\delta$ Scuti stars, RR Lyraes, Cepheids, eclipsing binaries, and long-period variables,
the superclasses, but also their subclasses, such as 
RRab, RRc, RRd, and RRe for RR Lyraes, and similarly for the other variable types.
The model trained using only the superclasses shows 99\% recall and precision,
while the model trained on all subclasses shows 87\% recall and precision.
We applied the trained model to the entire EROS-2 LMC database,
which contains about 29 million sources,
and found 117\,234 periodic variable candidates.
Out of these 117\,234 periodic variables, 55\,285 have not been discovered by
either OGLE or MACHO variability studies.
This set comprises 1906 $\delta$ Scuti stars, 6\,607 RR Lyraes,
638 Cepheids, 178 Type II Cepheids, 34\,562 eclipsing binaries, and 11\,394 long-period variables.
A catalog of these EROS-2 LMC periodic variable stars will be available online 
at \href{http://stardb.yonsei.ac.kr}{http://stardb.yonsei.ac.kr}
and at the CDS website (\href{http://vizier.u-strasbg.fr/viz-bin/VizieR}{http://vizier.u-strasbg.fr/viz-bin/VizieR}).}


\maketitle

\section{Introduction}

Studying periodic variable stars has
improved our understanding of the Universe for many decades.
For instance, Cepheid variables are one of the most important
variable types as a standard candle
for measuring extra-galactic distances \citep{Freedman2001ApJ, Riess2011ApJ}
because of their well-established period-luminosity relation \citep{Feast1997MNRAS, Storm2011AA},
which provided evidence for the expanding Universe \citep{Lemaitre1927ASSB, Hubble1931ApJ}.
RR Lyrae stars are useful for tracing 
the Galaxy formation history (e.g. see \citealt{Catelan2009ApSS} and references therein)
and for studying globular clusters \citep{Carretta2000ApJ}
and the Galactic structure \citep{Oort1975AA, Vivas2001ApJ}.
In addition, long-period variables such as Mira variables 
 show a period-luminosity
relation that can be used for measuring distances to some objects in the Galaxy,
such as globular clusters \citep{Feast1989MNRAS, Knapp2003AA}.
In brief, periodic variable stars 
are crucial for studying and understanding the Galaxy and the Universe.

The Exp$\acute{\text{e}}$rience pour la Recherche d'Objets Sombres (EROS)
is a wide-field sky survey for probing
dark matter in the Galactic halo by
detecting microlensing events (see \citealt{Tisserand2007AA} and 
references therein).
EROS monitored the Large/Small Magellanic Cloud (LMC/SMC),
the Galactic bulge and spiral arms for about seven years,
and was a major microlensing survey
together with the MACHO \citep{Alcock2000ApJ} and OGLE \citep{Udalski1997AcA}  microlensing surveys.
In addition to microlensing detections,
the EROS database is also useful
for detecting periodic variable stars
because of its well-sampled light curves
over a long period of observation,
relatively faint limiting magnitude of $\sim$20 in the EROS B band\footnote{EROS
B and R bands (i.e., $B_E$ and $R_E$) are not a standard astronomical B and R bands.
See Section \ref{sec:eros_database} for details.}, its wide field of view,
and two simultaneous passbands.

Previous studies have found
several types of variable stars in the EROS-2 database.
\citet{Beaulieu2001AA} discovered two variable stars
resembling Herbig Ae/Be or classical Be stars,
\citet{Tisserand2004AA} detected 
five R Coronae Borealis stars,
\citet{Marquette2009AA} discovered 185
new beat Cepheid variables,
\citet{Spano2011AA} reported forty-three thousand long-period variable candidates,
and \citet{Dubath2012IAUS} found about 300 variable
candidates from a subset of EROS-2 database but without identifying their variable types.
Although these works produced some variable star candidates of various types,
no study has searched 
the entire EROS-2 light curve database to classify different types of variable stars including 
$\delta$ Scuti stars, RR Lyraes, Cepheids, eclipsing binaries, and long-period variables.
For instance, \citet{Dubath2012IAUS} used a supervised machine-learning method
and multiple variability features to train a classification model,
but their training set is incomplete
because it consists of visually classified variables sorted into four classes
including periodic, small-amplitude,  semi-regular, and nonperiodic variables.
Thus the model cannot distinguish conventional types of astronomical variable stars
such as the types of variable stars listed above.

We initiated the  {\bf{E}}ROS-2 {\bf{P}}eri{\bf{O}}dic variable star {\bf{C}}lassification 
using mac{\bf{H}}ine learning (EPOCH) project\footnote{\href{http://stardb.yonsei.ac.kr}{http://stardb.yonsei.ac.kr}}
 that aims to detect periodic variables in
the  EROS-2 light curve database
to significantly increase the total number of known variable stars in the EROS-2 survey fields.
The EPOCH project is different from the previous studies 
because we used 1) the richest possible training set including multiple types of variable stars,
2) a few tens of variability features to separate variable stars from others, and
3) one of the most powerful supervised classification methods, random forest \citep{Breiman2001}. 
Random forest combines a large number of decision trees
to build a classification model and has successfully
solved many astronomical classification problems
(e.g. \citealt{Dubath2011MNRAS, Pichara2012MNRAS}).
Even though some of the above conditions
were fulfilled by previous work,
none of the works has fulfilled all of these conditions simultaneously.

In this paper,  we present the first results of the EPOCH project.
We also present periodic variable star candidates detected
from the EROS-2 LMC light curve database.
The EROS-2 database
is briefly introduced in Section \ref{sec:eros_database}.
Section \ref{sec:classifcation_for_periodic} presents a 
classification method including
1) the training set we used to build a classification model,
2) multiple variability features,
3) parameter optimization for the random forest model training, and
4) performance of a trained model.
We then show detection results of periodic variable stars
from the entire EROS-2 LMC database in Section \ref{sec:variable_candidate_selection}.
Section \ref{sec:summary} is a summary.

\section{EROS-2 database}
\label{sec:eros_database}

EROS-2 monitored the LMC/SMC, the Galactic center,
and the spiral arms during its operation from July 1996 to March 2003
using  the 1m Ritchey-Chr${\acute{\text{e}}}$tien telescope,
MARLY, at ESO (La Silla, Chile).
The telescope was equipped with two cameras,
one observing in the $B_E$ (420-720nm, blue) band, the other in the $R_E$ (620-920nm, red) band.
Each camera contained eight 2048 $\times$ 2048 CCD detectors in mosaics,
and had a field of view of 0.7$^{\circ}$$\times$1.4$^{\circ}$
in RA and Dec, respectively.
The pixel scale was 0.6$''$ and the typical seeing
at the site was 2$''$ FWHM \citep{Ansari2001ASPC, Tisserand2004AA}.

\begin{figure}
\begin{center}
       \includegraphics[width=0.5\textwidth]{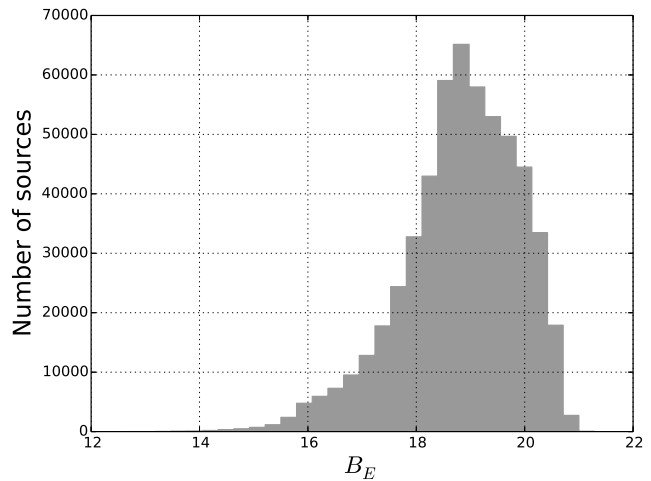}
\end{center}
    \caption{Histogram of EROS $B_E$ band magnitude of 
    about 550k field sources from the EROS-2 LMC fields.
    The limiting magnitude is around 20 $B_E$.}
    \label{fig:mag_histogram}
\end{figure}

The total number of sources in the EROS-2 database is
about 87 million consisting of
29 million in the LMC,
4 million in the SMC,
44 million in the Galactic bulge, and
10 million sources in the spiral arms.
Light curves obtained using the $B_E$ band
generally have better photometric accuracy
and also have more measurements
than $R_E$-band light curves \citep{Spano2011AA}.
Fig. \ref{fig:mag_histogram} shows
a histogram of the $B_E$ band magnitude
of about 550k sources in the EROS-2 LMC fields. 
The limiting magnitude of the field sources is $\sim$20 $B_E$.
The number of measurements for each light curve varies from field to field.
For instance, $B_E$-band light curves around the central region of the LMC
have 500 measurements on average,
while $B_E$-band light curves at the outer region
have  250 measurements on average.

\section{Classification of periodic variables}
\label{sec:classifcation_for_periodic}

In this section, we describe how we
classify periodic variable stars in the EROS-2 LMC light-curve database.
In Section \ref{sec:training_set}, we present a training set consisting of
EROS-2 variable stars selected by crossmatching with
the known OGLE and MACHO variables.
Section \ref{sec:reduction} explains how we refined  the light curve
(e.g., cleaning spurious data points)
before we extracted 22 variability features
that are described in Section \ref{sec:features}.
We then present the random forest classification algorithm
and summarize its performance in Section \ref{sec:classification_model}.

\subsection{Training set}
\label{sec:training_set}

The classification quality 
of supervised machine-learning methods relies heavily on the training set.
To build the richest possible training set,
we therefore used  the OGLE LMC variable star catalogs 
(\citealt{Soszynski2008AcA, Soszynski2008AcA2, Soszynski2009AcA, 
Soszynski2009AcA2, Poleski2010AcA, Graczyk2011AcA}),
which are the most up-to-date and complete catalogs of periodic variable stars
over the EROS-2 fields.
Based on these catalogs, we compiled a list 
of periodic variable stars including
$\delta$ Scuti stars, RR Lyraes, Cepheids, Type II Cepheids, 
eclipsing binaries, and long-period variables
(DSCTs, RRLs, CEPHs, T2CEPHs, EBs and LPVs, respectively, Table \ref{tab:acronym}).
We adopted the OGLE classification scheme for variable subclasses 
(e.g. RRL ab, c, d, and e type; see Table \ref{tab:training_set} for the full list).
We added 982 blue variables detected from the MACHO database \citep{Keller2002AJ}.
Many of these blue variables are probably Be stars
since their variability characteristics are consistent with those of Be stars \citep{Keller2002AJ}.
We also added 565 quasi-stellar objects (QSOs) to the list that were selected 
using the OGLE LMC database \citep{KozlowskiarXiv2013}.
Although we do not aim to classify either
blue variables (BVs) or QSOs from the EROS-2 database,
including these pseudo- or non-periodic variables in the list
is important to exclude these variable types
during the selection of periodic variable candidates. 
There  are many BVs in the LMC, some of which show periodic patterns in their light curves.
We did not include variable candidates from 
\citet{Marquette2009AA},  \citet{Spano2011AA}, and \citet{Dubath2012IAUS},
because they have too few variable candidates of each type.
More importantly, using variable sources mainly from a single survey (i.e., OGLE)
helps maintain consistency of the training set.
The total number of sources in the compiled catalog is 149\,388.

\renewcommand{\arraystretch}{1.2}
\begin{table}
\small
\begin{center}
\caption{Acronyms of each variable type \label{tab:acronym}}
\begin{tabular}{cc}
\hline\hline
Variable type & Acronym \\
\hline
$\delta$ Scuti & DSCT \\
RR Lyrae & RRL \\
Cepheid & CEPH \\
Type II Cepheid & T2CEPH \\
Eclipsing binary & EB \\
long-period variable & LPV \\
Blue variable & BV \\
\hline
\end{tabular}
\end{center}
\end{table}

\begin{figure}
\begin{center}
       \includegraphics[width=0.5\textwidth]{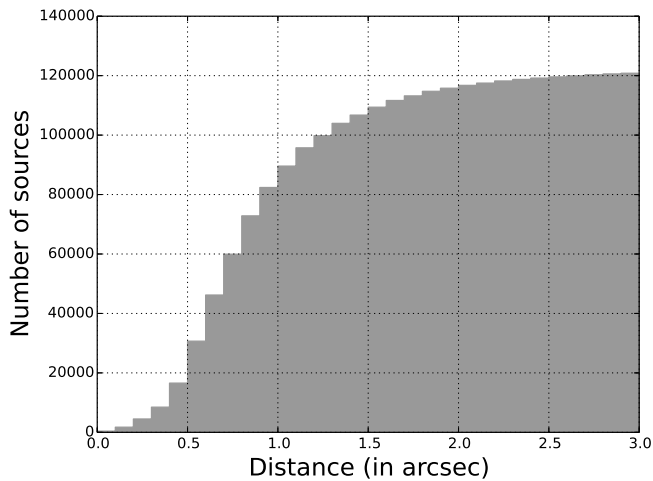}
\end{center}
    \caption{Cumulative histogram of distances to the EROS-2 counterparts  crossmatched 
    with the previously known variables.
    The largest search radius for crossmatching is 3$''$.
    Most of the variables are crossmatched
    within a 1.5$''$ search radius.}
    \label{fig:crossmatching_distance}
\end{figure}

We then crossmatched these variables with the
entire EROS-2 LMC database with a 3$''$ search radius
that yielded 120\,825 counterparts,
119\,480 of which are OGLE periodic variable stars
and the remaining 1\,345 sources are QSOs or BVs.
Fig. \ref{fig:crossmatching_distance} shows
a cumulative histogram of distances to the counterparts.
To minimize the number of  incorrect crossmatches, we only
selected sources among these counterparts
where the distance between the nearest neighbor 
and second-nearest neighbor is larger than 3$''$. 
After removing these objects, 97\,912 EROS-2 sources remained.

\begin{figure*}
\begin{center}
       \includegraphics[width=1\textwidth]{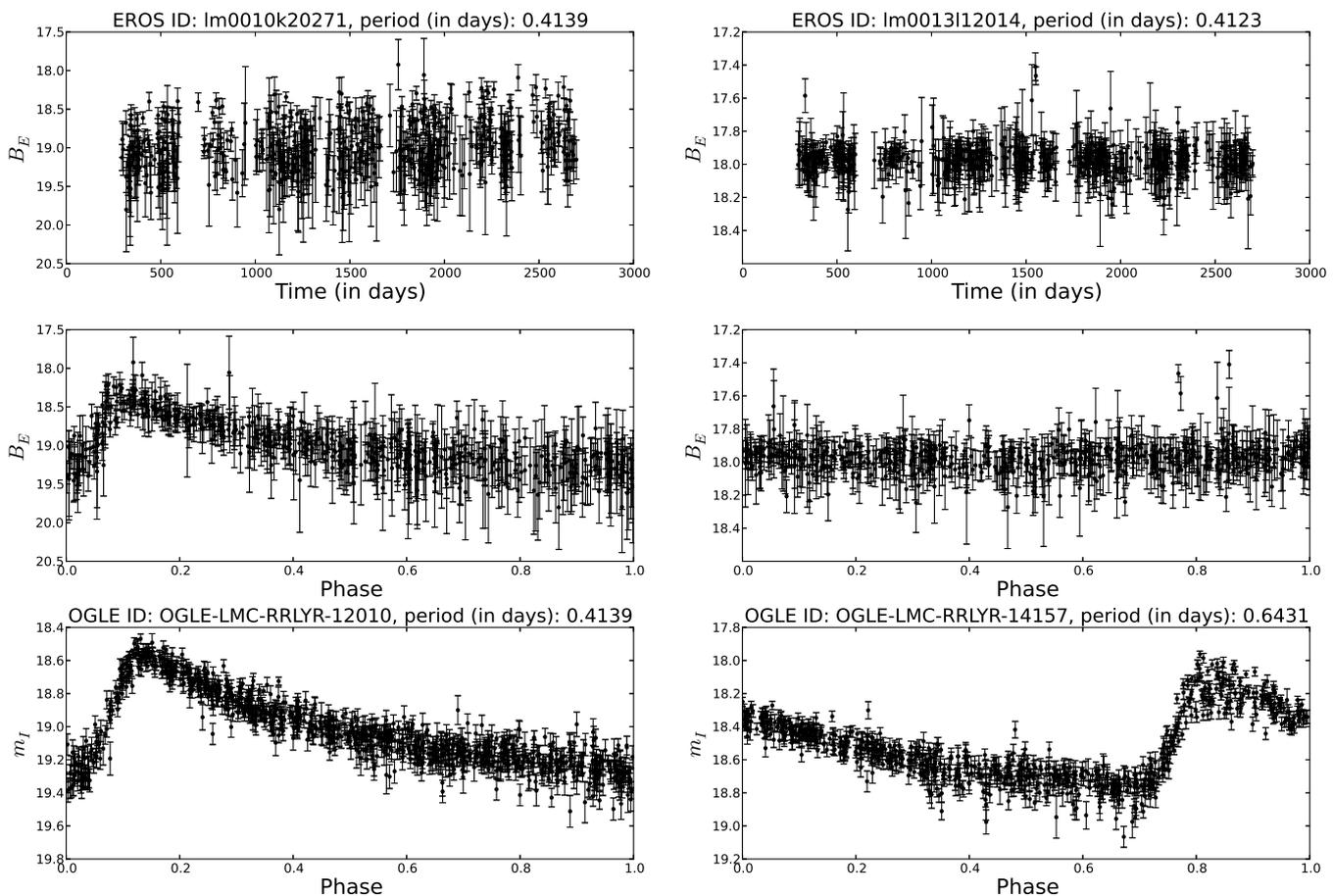}
\end{center}
    \caption{EROS-2 light curves (top and middle) of two variable stars crossmatched with the OGLE RRLs.
    Top: raw light curves, middle: phase-folded light curves,
    left: one RRL variable included in the training set, and right: another RRL variable excluded from the training set.
    The x-axis of raw light curves is in the unit of Heliocentric Julian Date (HJD) - 2\,450\,000.
    We use this unit as the unit of x-axis for a raw light curve throughout this paper.
    Periods shown in the top panels are derived using the Lomb-Scargle algorithm \citep{Lomb1976ApSS}.
    In the bottom two panels, we show the phase-folded OGLE light curves of the same RRLs. 
    The periods of these light curves are taken from \citet{Soszynski2009AcA2}.
    Although the OGLE RRL at the bottom-right panel shows clear periodic signals,
    the crossmatched EROS-2 counterpart's light curve  (top right and middle right) does not show any variability,
    which is associated with blending with a nearby star.
    See text for details.
}
    \label{fig:rrl_visual}
\end{figure*}

Even if the known OGLE variables
show clear variability in their light curves,
it is possible that the crossmatched
EROS-2 sources might not show clear variability
because of the photometric uncertainty, blending with other sources,
different sampling rate, or incorrect crossmatching.
To exclude such light curves from a training set, 
one of us (DWK) visually examined raw- and phase-folded light curves
of all 97\,912 EROS-2 counterparts.
During the visual examination,
we removed the light curves without variability or periodicity.
Such a pure training set is important 
because the goal of this work is 
to select periodic variable candidates from the entire EROS-2
database, which consists of several tens of millions of light curves.
Therefore minimizing false positives mainly consisting of non-variables
is the most critical task,
which is facilitated by having a high-quality training set.
Fig. \ref{fig:rrl_visual} shows
example light-curves of two RRL variables.
We retained light curves with clear variability in the EROS-2 light curves,
as shown in the left panels.
We visually examined an EROS-2 reference image for the source
shown in the right panel of Fig. \ref{fig:rrl_visual}
and found that the source is blended with a nearby bright star
identified as a RRL variable by the OGLE.
This particular case shows that
the OGLE differential photometry (the DIA, \citealt{Alard1998ApJ})
is more efficient than the fixed-position photometry 
of the EROS (the PEIDA, \citealt{Ansari1996VA}) 
to identify variable stars in crowded fields.
After this visual removal, we had a set of 27\,892 periodic variables.

We finally added non-variable sources
to this set of visually cleaned light curves by 
randomly selecting about 5\,000 
light curves that are spread over the whole
EROS-2 LMC fields and are not in the 120\,825 EROS counterparts 
mentioned in the previous paragraph.
Adding non-variables to the training set is critical
when selecting variable candidates
from the large EROS-2 database
because the majority of light curves are non-variables.
Therefore a classification model has to be trained 
in such a way that it excludes non-variables during the variable-star selection.
To balance the number of light curves
through the EROS-2 magnitude range, we selected
about 800 non-variable light curves for each of the magnitude ranges,
13-16, 16-17, 17-18, 18-19, 19-20, and 20-21  $B_E$.
If  we had randomly selected light curves without the regular magnitude range, 
there would have been substantially fewer light curves at the brighter end 
than light curves at the fainter end,
which is undesirable because then a trained model
might be inefficient in excluding possible non-variables at the bright end.
We also visually examined these non-variable light curves
and excluded light curves with variability.
After adding these non-variables, the final number of light curves in the training set is  32\,683.

Without the visual removal of non-variable sources mentioned in the previous paragraph, 
we observed a significant drop of classification performance,
in particular a performance decreased  by $\sim$20\% for non-variables
(see Section \ref{sec:training_classificatio_model}
for the definition of classification performance).
This decrease could cause many
misclassifications of non-variables into variables,
which is not desired because most of the light curves from the 
EROS-2 database are expected to be non-variables.
Nevertheless, even though we did not intentionally remove weak-variability sources,
our visual method might result 
in the rejection of some true variables at the end
because of the low signal-to-noise ratio
of the variables.\footnote{Signal-to-noise
ratio here means the highest amplitude versus noise of a light curve.}
In Section \ref{sec:sanity},
we show test results that alleviate these concerns for classifying weak-variability sources.

Table \ref{tab:training_set} shows the number of variables
of each type 1) after the crossmatching, 
2) after the visual examination, and
3) after the light-curve refining explained in Section \ref{sec:reduction}.
Although we maintained most of the subclasses of variable types
defined from the OGLE catalogs, we merged some subclasses 
that do not have many sources.
For instance, the subclass ``Other'' of CEPHs includes 
second-overtone (i.e., 2O), double-mode F/1O (i.e., fundamental-mode and first-overtone), 
double-mode 1O/2O, 
double-mode 1O/3O (i.e., third-overtone) and triple-mode classical CEPHs.
One of the EB subclasses, ``Other'',
also includes several types of eclipsing binaries.
We also did not separate subclasses of T2CEPHs
and DSCTs because of insufficient numbers.

\renewcommand{\arraystretch}{1.2}
\begin{table*}
\small
\begin{center}
\caption{Number of remaining variables\label{tab:training_set}
after crossmatching, visual removal, and light-curve refining}
\begin{tabular}{cccccc}
\hline\hline
Superclass & Subclass & After crossmatching &  After visual removal & After refining & Note \\
& & (number) & (number (\%$^a$)) & (number (\%$^a$)) &  \\
\hline
DSCT & & 1155 & 511 (44) & 511 (44) & \\
CEPH & & & & \\
		& fundamental (F) & 1275 & 1275 (100) & 1249 (98) & \\
		& first overtone (1O) & 871 & 844 (97) & 834 (96) & \\
		& Other & 168 & 162  (96) & 160 (95) & \\
RRL & & & &  & \\
		& ab & 8721 & 3434 (39) & 3276 (38) & \\
		& c & 2461 & 714 (29) & 637 (26) & \\
		& d & 511 & 179 (35) & 172 (34) & \\
		& e & 603 & 143  (24) & 141 (23) & \\
EB & & & &  & \\
		& EC & 605 & 445 (74)  & 418 (69) & contact \\
		& ED & 8615 & 1644 (19) & 1592 (19) &  detached \\
		& ESD & 3409 & 1150 (34) & 1109 (33) & semi-detached \\
		& ED+ESD & 870 & 146 (17) & 142 (16) & detached/semi-detached \\
		& Other & 254 & 153 (60)  & 146 (58) &  \\
LPV & & & &  & \\
		& Mira AGB C & 785 & 754 (96)  & 567 (72) &  carbon-rich \\
		& Mira AGB O & 329 & 329  (100) & 324 (99) &  oxygen-rich \\
		& OSARG AGB C & 2119 & 1416 (67)  & 1135 (54) & carbon-rich \\
		& OSARG AGB O & 22670 & 3368  (15) & 3187 (14) & oxygen-rich \\
		& OSARG RGB C & 264 & 57 (22) & 46 (17) & carbon-rich \\
		& OSARG RGB O & 32648 & 2046 (6) & 1888 (6)& oxygen-rich \\
		& SRV AGB C & 3889 & 3819  (98) & 2507 (65) & carbon-rich \\
		& SRV AGB O & 4482 & 4206 (94) & 3672 (82) & oxygen-rich \\
T2CEPH & & 123 & 121 (98) & 112 (91) & \\
BV & & 829 & 796 (96)  & 725 (88) &  \\
QSO & & 256 & 180  (78) & 161 (63) &  \\
NonVar & & & 4791 & 3681 (77) &  \\
\hline
Total & & 97912 &  32683 (33) & 28392 (29) & \\
\end{tabular}
\end{center}
${}^a$ ratio to the number of variables after crossmatching.
\end{table*}

\begin{figure}
\begin{center}
       \includegraphics[width=0.5\textwidth]{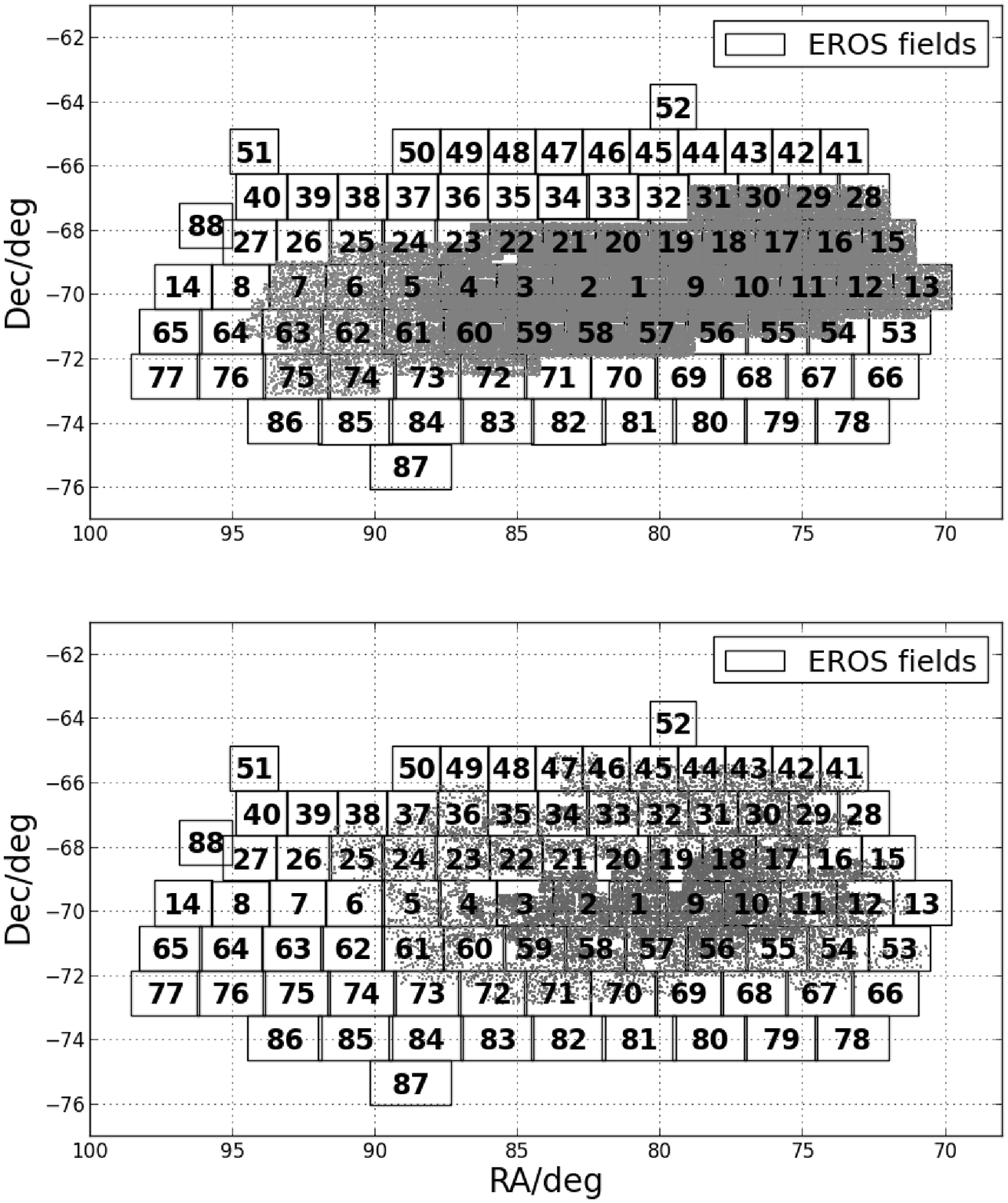}
\end{center}
    \caption{EROS-2 sources  (gray dots) crossmatched with 
    the known OGLE (top) and MACHO (bottom) variable stars
    in the EROS-2 LMC fields.
    The rectangles in each panel show the EROS-2 fields.
    The integer numbers in the rectangles are the EROS-2 field IDs.
    The x- and y-axis is RA and Dec is in J2000.
    }
    \label{fig:sky_coverage}
\end{figure}

The top panel of Fig. \ref{fig:sky_coverage} shows 
the OGLE counterparts before the visual removal
superposed on the coverage of the EROS-2 LMC fields.
The counterparts are only spread over the central region of the EROS-2 fields.
About half of the region
is not overlapped by the OGLE counterparts.
In addition, we show the crossmatched counterparts 
with the MACHO variables in the
bottom panel of Fig. \ref{fig:sky_coverage}.
We used the list of MACHO variables from \citet{Kim2011ApJ},
which contains about 19k sources including RRLs, CEPHs, EBs, LPVs, and BVs.
The MACHO variable study \citep{Alcock2001} provides
one of the richest variable catalogs 
in terms of the number of variables and variable types
along with the OGLE variable studies.
As the panel shows, the MACHO variables
also do not cover the entire EROS-2 fields.
Thus it is clear that there are numerous variable stars
that have never been detected by either the OGLE or the MACHO survey.

Note that the training set explained in this section
is mainly based on the OGLE variable stars
that are spread across the central region of the LMC.
Thus it is possible that a classification model
built on the training set might be biased
to characteristics of the light curves in the central LMC area 
such as magnitudes, amplitudes, colors, and photometric uncertainties.
Nevertheless, training individual classification models
adapted to each EROS-2 field's characteristics  is beyond the scope of this paper.

\subsection{Refining light curve} 
\label{sec:reduction}

To extract variability features,
we used EROS-2 $B_E$-band light curves,
which have better photometric accuracy
and also more data points than $R_E$-band light curves.
This is because the red camera was subject to more technical problems than the blue one during all campaigns.
Before extracting variability features from the training set light-curves explained in the previous section,
we refined each light curve as follows:

\begin{itemize}

\item We removed measurements with
magnitudes higher than 22 or 
with photometric uncertainties higher than 1 magnitude.
This cut removes many measurements of a light curve
at the faint-end magnitude 
because such a light curve mainly consists of
99.999 magnitude values or 9.999 photometric uncertainty values,
which indicates that the measurements are unreliable.

\item We removed measurements by 3$\sigma$ clipping (no iteration) about the mean, 
where $\sigma$ is a standard deviation of the light curve.
This cut is used to remove fluctuating data points
that are occasionally caused by inaccurate photometric measurements.
Nevertheless, the cut typically removes less than 5\% of
the data points, which does not significantly alter light curves.

\end{itemize}

We then excluded light curves from the training set if

\begin{itemize}

\item there were fewer than 100 measurements.

\item the estimated period was spurious, related to
the solar day, the moon phase, the sidereal months, and their multiples.
We found spurious periods by examining 
a scatter plot of periods versus period S/N (e.g. see the top-left panel
in Fig. \ref{fig:features_plot})  and a period histogram.

\end{itemize}

{\noindent}This reduced the number of light curves reduces from 32\,683 to 28\,392.
Most of the excluded light curves were removed by  the period criterion.
This set of 28\,392 light curves is the final training set.

\subsection{Variability features}
\label{sec:features}

To identify  variability,
we used multiple features extracted from the light curves
instead of the light curve themselves as the 
basis for classification.
We tested more than 30 features and selected
the best 22 features based on their ``variable importance'',
estimated using the random forest algorithm explained in Section \ref{sec:randomforest}.
We found empirically that using more or fewer features than 22
did not improve classification results.
\citet{Dubath2011MNRAS} also observed that
using a large number of features, including inessential or even correlated features
did not affect random forest classification results much.

\begin{figure}
\begin{center}
       \includegraphics[width=0.5\textwidth]{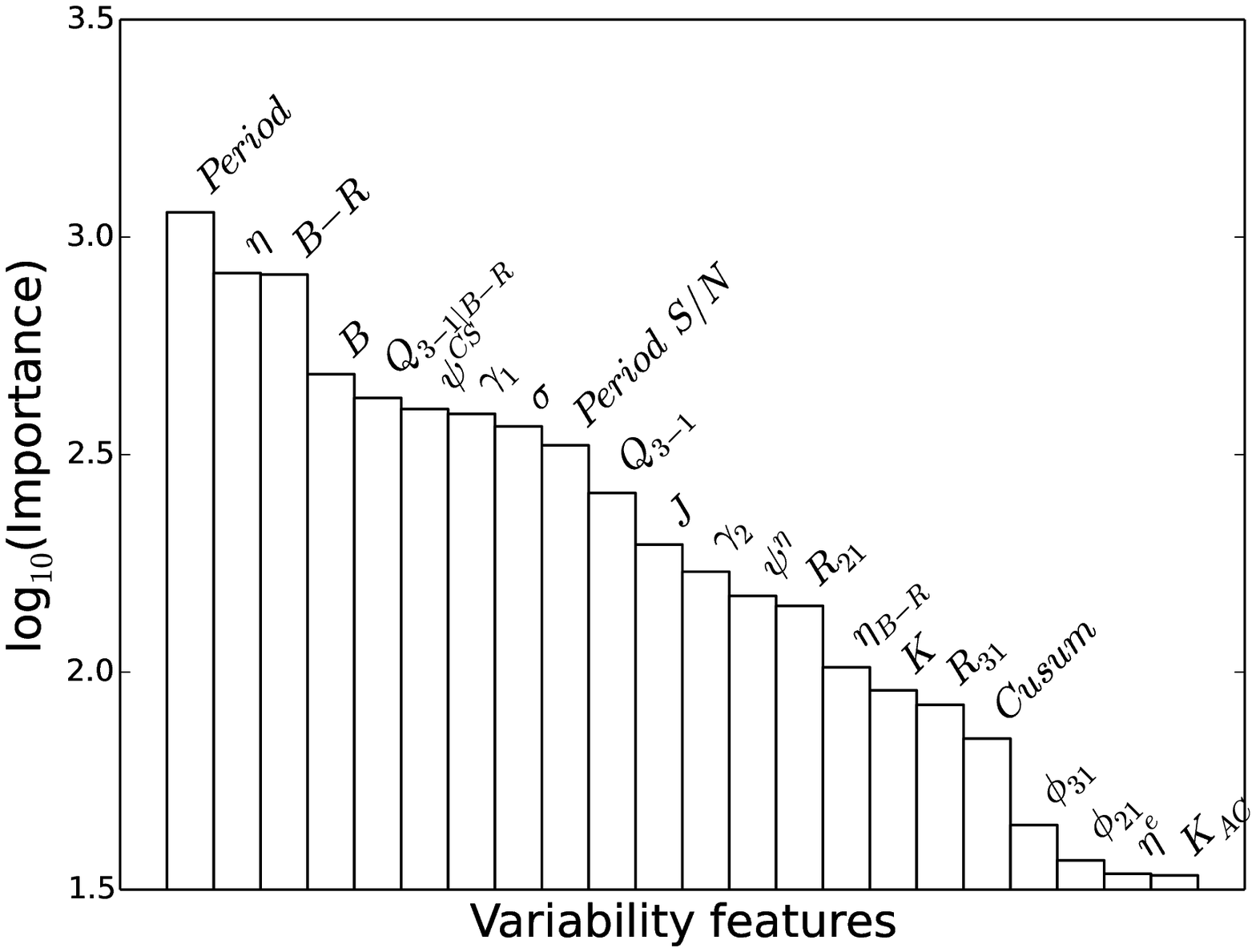}
\end{center}
    \caption{Variable importance of the variability features as determined 
    by the random forest algorithm considering
    the variable subclasses.
    Higher importance features are more relevant for classification
    than lower importance features. A brief summary of these features
    is given in Table \ref{tab:time_variability_features}.}
    \label{fig:variable_importance}
\end{figure}

A brief description of each feature is given in
Table \ref{tab:time_variability_features}.
Fig. \ref{fig:variable_importance} shows
the estimated variable importance for these 22 features.
For details of the definitions of the indices, see the appendix.

\renewcommand{\arraystretch}{1.2}
\begin{table*}
\small
\begin{center}
\caption{Variability features \label{tab:time_variability_features}}
\begin{tabular}{cl}
\hline\hline
Feature & A brief description. For details, see the appendix. \\
\hline
Period & Period derived by the Lomb-Scargle algorithm \\
Period S/N & Signal-to-noise ratio of the derived period \\
$\eta$ & Ratio of the mean of square of successive differences to the variance of data points \\
$\eta_{B-R}$ & $\eta$ over a $B_E$ - $R_E$ light curve \\
$\psi^{\eta}$ & $\eta$ over a phase-folded light curve \\
$\eta^e$ & modified $\eta$ \\
$B_E$ & Mean magnitude of $B_E$ band\\
$B_E - R_E$ & Color. Difference of mean magnitude of  $B_E$ and $R_E$ bands\\
$Q_{3-1}$ & 3$^{rd}$ quartile (75\%) - 1$^{st}$ quartile (25\%) \\
$Q_{3-1|B-R}$ & 3$^{rd}$ quartile (75\%) - 1$^{st}$ quartile (25\%) over a $B_E$ - $R_E$ light curve\\
Cusum & Range of cumulative sum \\
$\psi^{CS}$ & Cusum over a phase-folded light curve \\
$\sigma$ & Standard deviation \\
$\gamma_1$  & Skewness \\
$\gamma_2$ & Kurtosis \\
$J$ & Stetson $J$ index \\
$K$ & Stetson $K$ index \\
$K_{AC}$ & Stetson $K$ index derived over an autocorrelation function of a light curve \\
$R_{21}$ & 2$^{nd}$ to 1$^{st}$ amplitude ratio derived using the Fourier decomposition \\
$R_{31}$ & 3$^{rd}$ to 1$^{st}$ amplitude ratio derived using the Fourier decomposition \\
$\phi_{21}$ & Difference between 2$^{nd}$ and 1$^{st}$ phase  derived using the Fourier decomposition \\
$\phi_{31}$ & Difference between 3$^{rd}$ and 1$^{st}$ phase derived using the Fourier decomposition \\
\hline
\end{tabular}
\end{center}
Notes: We used EROS $B_E$-band light curves (either raw- or phase-folded) to derive these features,
except for $J$, $Q_{3-1|B-R}$ and $\eta_{B-R}$,
which needs both $B_E$- and $R_E$-band light curves.
See the appendix for details.
\end{table*}

\subsection{Random forest classification model}
\label{sec:classification_model}

\subsubsection{Architecture and training}
\label{sec:randomforest}

To train a classification model
using the variability features extracted from the training set light-curves,
we used the random forest classification method \citep{Breiman2001}.
This has been successfully  applied for many astronomical classification and
regression problems (e.g. 
\citealt{Carliles2010ApJ, Dubath2011MNRAS, Richards2011ApJ, Pichara2012MNRAS}).
Random forest uses multiple decision trees \citep{Quinlan1993} and 
the bagging \citep{Breiman1996} to train a classification model.
Each decision tree is trained using a subset of features that are
randomly selected from all features (here the 22 variability features mentioned in the previous section).
Random forest then chooses a class for a sample that has the most
votes of the all decision trees.
In principle, random forest does not need to do cross-validation
to estimate the classification error because each tree
is constructed based on the bootstrap sample from the training set,
where two thirds of the samples are used
to train the tree and the remaining one third of samples, 
which are out-of-bag (oob) samples, 
is used to estimate an error, which is an oob error.
The process of constructing each decision tree is as follows:

\begin{enumerate}

\item Generate  $N$ light curves randomly selected from the training set with replacement,
where $N$ is the total number of light curves in the training set, which is called bootstrap aggregating (bagging).
This sample is used to train the tree.

\item At each node, randomly select $m$  features from the
all features, where $m<M$. $M$ is the total number of features.
The $m$ value is fixed during the training.

\item These randomly selected $m$ features are used to split the node. 
At each node, every possible split is tested,
and then a feature for the best split is chosen.
The tree is grown to its full extent without pruning.
Each leaf node of the tree returns a single sample.

\item Repeat $t$ times the above processes, where $t$ is
the number of decision trees.

\end{enumerate}

{\noindent}While training a random forest model,
one needs to set two free parameters, $t$ and $m$, 
which is the number of decision trees and the number of features, respectively.
We briefly explain the parameter selection in the following section.

We can use random forest
to estimate the variable importance relevant to classification as follows:

\begin{enumerate}

\item Train a normal random forest model, $RF_{normal}$.

\item Train another random forest model 
after randomly permuting values of a feature among the training set, 
$RF_{permute}$.

\item During the training processes, calculate differences in oob error
between the trees from the $RF_{normal}$ and the $RF_{permute}$.
Average the differences over all trees and
normalize the average differences by its standard deviation derived from the differences.
This averaged and normalized difference is a measure of the importance of the feature.

\end{enumerate}

{\noindent} Thus stronger features generally have higher variable importance values.
The derived variable importance of the 22 variability
features is given in Fig. \ref{fig:variable_importance}.

Random forest results can be used to estimate class probabilities
for each light curve based on a simple voting strategy,

\begin{equation}
P_{class}(i) = \frac{n_{class}(i)} {N},
\end{equation}

{\noindent}where $P_{class}$ is the probability of a $class$ for a sample, 
$n_{class}$ is the number of
trees that identify the sample as $class$, and $N$ is
the total number of trees.
For each light curve, we accepted the class corresponding to the highest probability
among probabilities of all classes. 
We did not modify the probability according to class population in the training set
since it is not the true population.

\subsubsection{Training a classification model and performance optimization}
\label{sec:training_classificatio_model}

Although random forest can internally estimate
the classification error, as mentioned in the previous section (oob error),
we separated the training set mentioned in Section \ref{sec:training_set}
into two sets of samples to 1) measure performance, and 
2) find the best $t$ (the number of trees) and $m$ (the number of features).
The samples consist of 
50\% of the training set ($\sim$14\,000 light curves) to train the model and
the remaining 50\% to estimate the trained model's classification performance.
We call these two set $S_1$ and $S_2$.
We assessed the classification performance using 
a weighted average of recall and precision by class size in the training set.
The recall and precision are defined as

\begin{eqnarray}
\begin{array}{l}
\displaystyle
{\rm{recall}} = \frac{N_{TP}}{N_{TP} + N_{FN}}, \,\,{\rm{precision}} = \frac{N_{TP}}{N_{TP} + N_{FP}},
\end{array}
\end{eqnarray}

\begin{figure}
\begin{center}
       \includegraphics[width=0.5\textwidth]{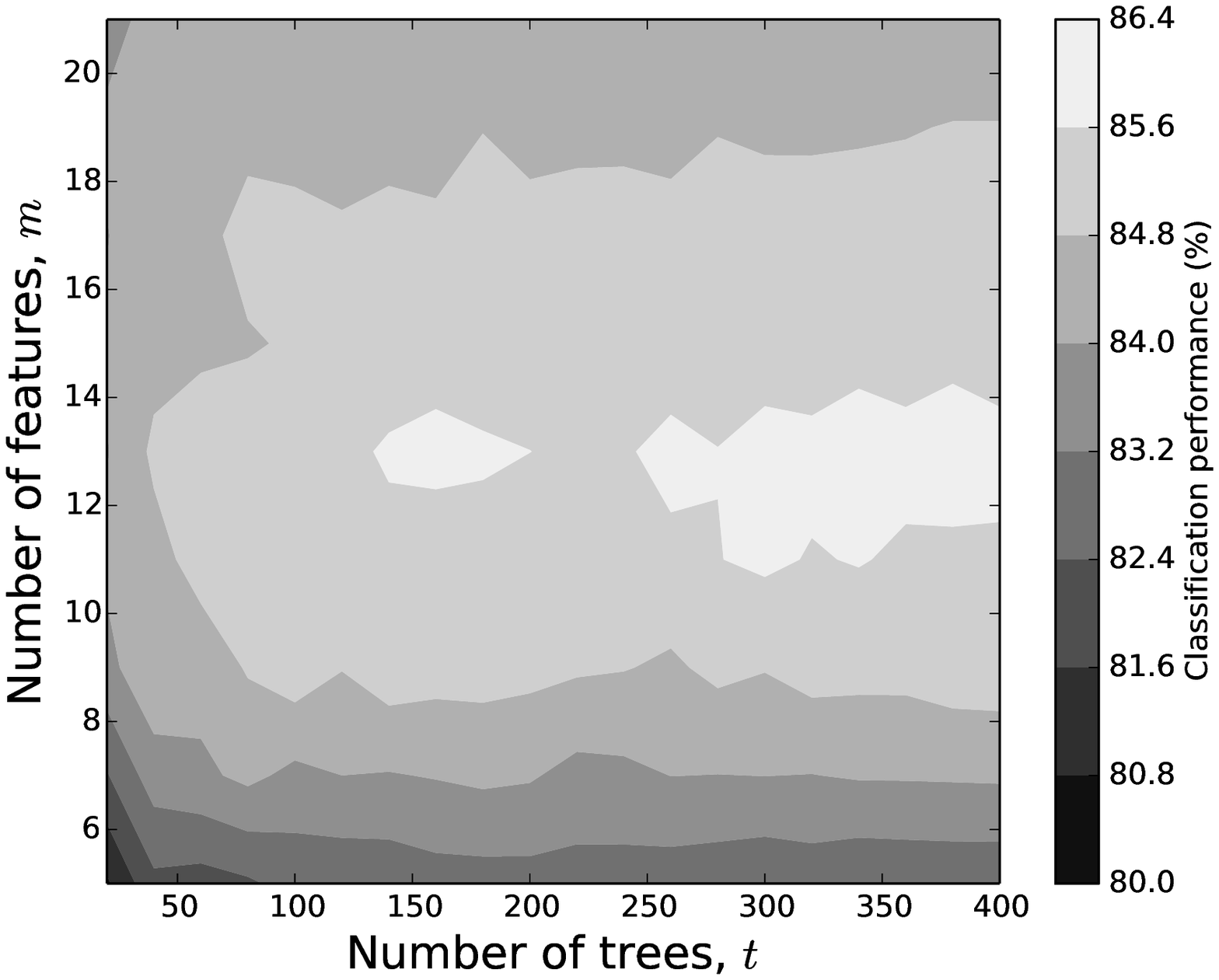}
\end{center}
    \caption{Classification performance according to
    the number of trees, $t$, and the number of features, $m$.
    Performance degrades when m $>$ 14, which might be
    caused by overfitting of a classification model.}
    \label{fig:grid_search_performance}
\end{figure}

{\noindent}where $N_{TP}$ is the number of the true positives,
$N_{FP}$ is the number of the false positives, and
$N_{FN}$ is the number of the false negatives.
To select the best $t$ and $m$,
we trained a random forest model for all possible combinations  of $t$  and $m$
with $20 \le t \le 400$ and $5 \le m \le 21$
with a step size of 20 and 2, respectively.
We then applied each trained model to $S_2$
to measure the model's performance.
Finally, we chose the $t$ and $m$ with the best performance.
Fig. \ref{fig:grid_search_performance} shows the classification performance
for each combination of $t$ and $m$. 
The classification performance through this parameter space
is relatively stable, as shown in the figure.

Tables \ref{tab:mixedLibraryConfMatrix} and \ref{tab:LibraryConfMatrix}
show the performance of the trained models
using the optimized parameters, $t$ and $m$.
In Table \ref{tab:mixedLibraryConfMatrix}, we present the confusion matrix
of the model trained with only superclasses (i.e., without considering subclasses of variable types).
We see that recall and precision for
each superclass is higher than 95\%,
and average recall and precision is about 99\%.
Only the T2CEPH class shows low recall and precision.
This is because 1) the number of variables in
this class is substantially lower than that of the other classes, and
2) the variability characteristics of T2CEPHs are
similar to other classes (i.e., sinusoidal pattern of the light curves).
Table \ref{tab:LibraryConfMatrix} shows
the recall and precision of the model trained with the subclasses\footnote{Due
to the length of the confusion matrix, we do not show it but just
show the recall and precision for each class. The confusion matrix
is available as supplementary material.}.
The average recall and precision drop to 87\%,
not because of the misclassification
between superclasses,
but mainly because of the misclassification of subclasses within a superclass.
This is not surprising since
the classification of each subclass within 
a same superclass is harder than the
separation of superclasses.
It is also possible that the OGLE
classification is not sufficiently accurate, 
especially for EBs and LPVs, 
as the  OGLE papers  \citet{Soszynski2009AcA} and \citet{Graczyk2011AcA} 
indicate.

\begin{figure*}
\begin{center}
       \includegraphics[width=1.0\textwidth]{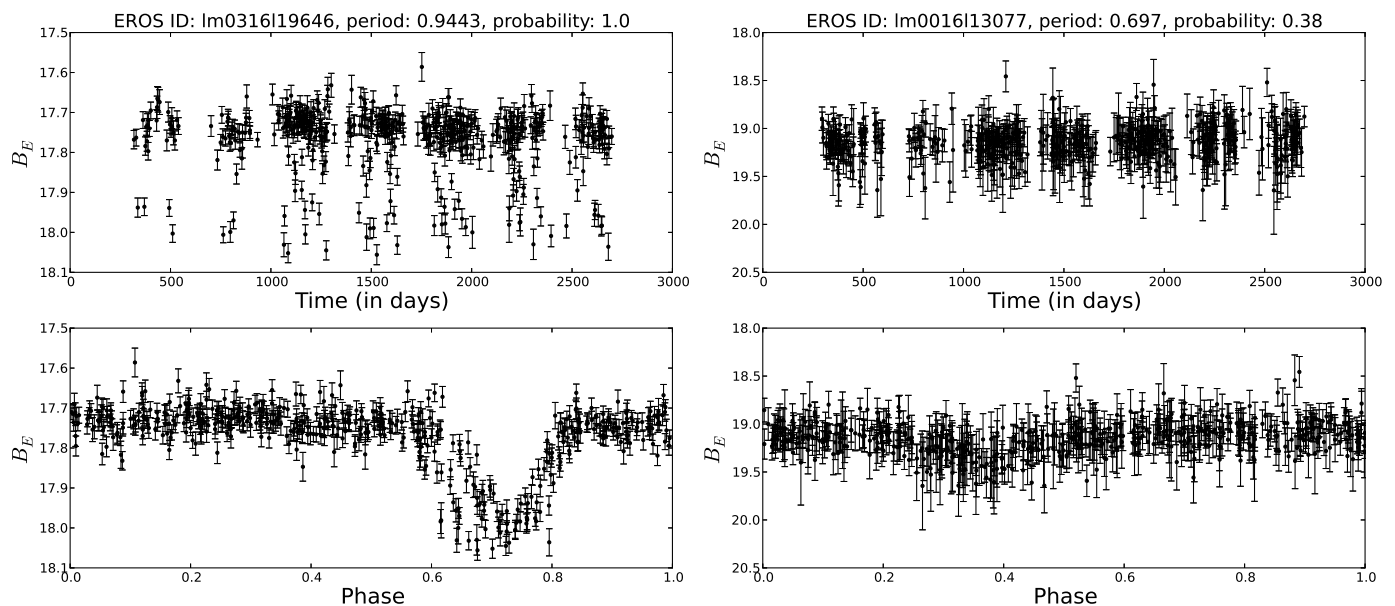}
\end{center}
    \caption{Example light curves of two EB variables 
    with a high (left panel) and low probability (right panel).
    The two left (right) panels are light curves of a same star.
    The top panels are raw light curves
    and the bottom panels are phase-folded light curves.
    The light curve with a high probability
    shows much stronger and clearer transit signals
    than the other light curve.}
    \label{fiig:high_low_probability_light_curve}
\end{figure*}

\renewcommand{\arraystretch}{1.2}
\begin{table*}
\small
\begin{center}
\caption{Confusion matrix of the random forest model without considering subclasses$^a$  \label{tab:mixedLibraryConfMatrix}}
\begin{tabular}{c|ccccccccc|c}
\hline\hline
Class & DSCT & RRL & CEPH & T2CEPH & EB & LPV & BV & QSO & NonVar & Recall\\
 &  &  &  &  &  &  &  &  &  & (\%)\\
\hline
DSCT & \bf 253 & 1 &  &  &  &  &  &  & 1 & 99.2 \\
RRL & 3 & \bf 2088 &  4 & 2 &  16 &  &  &  &  & 98.8 \\
CEPH &  & 10 & \bf 1094 &  3  &  12 &  & 1 &  & 1 & 97.6 \\
T2CEPH &   &  &  8 & \bf 37 &  10 & 1 &  &  &  & 66.1 \\
EB &  1 & 11 & 6 &    & \bf 1652 &  20 &  &  & 14 & 96.9 \\
LPV &   &  &  &   1 & 6 & \bf 6642 & 1 & 1 & 12 & 99.7 \\
BV &   &  &  &  & 2 &  & \bf 353 & 1 & 6 & 97.5 \\
QSO &   &  &  &   &  & 1 & 2 & \bf 71 & 7 &  87.7 \\
NonVar &  2 & 1 &  &   & 6 & 12 & 5 & 4 & \bf 1812 &  98.4 \\
\hline
Precision & 97.7 & 99.0 & 98.4 & 86.0 & 96.9 & 99.5 & 97.5 & 92.2  & 97.8  & 98.6, 98.6$^b$\\
(\%) &  &  &  &  & &  &  &   &   & \\
\end{tabular}
\end{center}
$^a$ Each row is a true class and each column is a predicted class. 
{\bf{Bold}} numbers show the true positives.\\
$^b$ Average weighted recall and precision by the proportion of classes in the training samples.
\end{table*}

\renewcommand{\arraystretch}{1.2}
\begin{table*}
\small
\begin{center}
\caption{Recall and precision of the random forest model considering subclasses\label{tab:LibraryConfMatrix}}
\begin{tabular}{cccc}
\hline\hline
Class & Subclass & Recall & Precision \\
& & (\%) & (\%) \\
\hline
DSCT & & 97.6 & 98.0\\
CEPH & & & \\
  & F & 97.3 & 97.7 \\
  & 1O & 91.1 &91.6 \\
  & Other & 71.3 & 73.1 \\
RRL & & & \\
  & ab & 99.6 & 99.0 \\
  & c & 95.9 & 87.4 \\
  & d & 65.1 & 93.3 \\
  & e & 83.1  & 85.5 \\
EB & & & \\
  & EC & 52.6 & 70.1 \\
  & ED & 88.9  & 79.9 \\
  & ESD & 73.0 & 65.0 \\
  & ED+ESD & 5.6 & 50.0 \\
  & Other & 11.0 & 44.4 \\
LPV & & & \\
  & Mira AGB C & 88.3 & 86.5 \\
  & Mira AGB O & 82.1 & 86.9 \\
  & OSARG AGB C & 57.0 & 71.9 \\
  & OSARG AGB O & 80.8 & 76.1 \\
  & OSARG RGB C & 13.0 & 30.0 \\
  & OSARG RGB O & 84.4 & 83.4 \\
  & SRV AGB C & 88.2 & 85.3 \\
  & SRV AGB O & 84.4 & 84.9 \\
T2CEPH & & 71.4 & 88.9 \\  
BV & & 96.1 & 97.2 \\
QSO & & 92.5 & 91.4 \\
NonVar & & 98.2 & 97.6 \\
\hline
Average recall and precision$^a$ & & 86.7 & 86.3  \\
\end{tabular}
\end{center}
$^a$  Average weighted by the proportion of classes in the training samples.
\end{table*}

\begin{figure}
\begin{center}
       \includegraphics[width=0.5\textwidth]{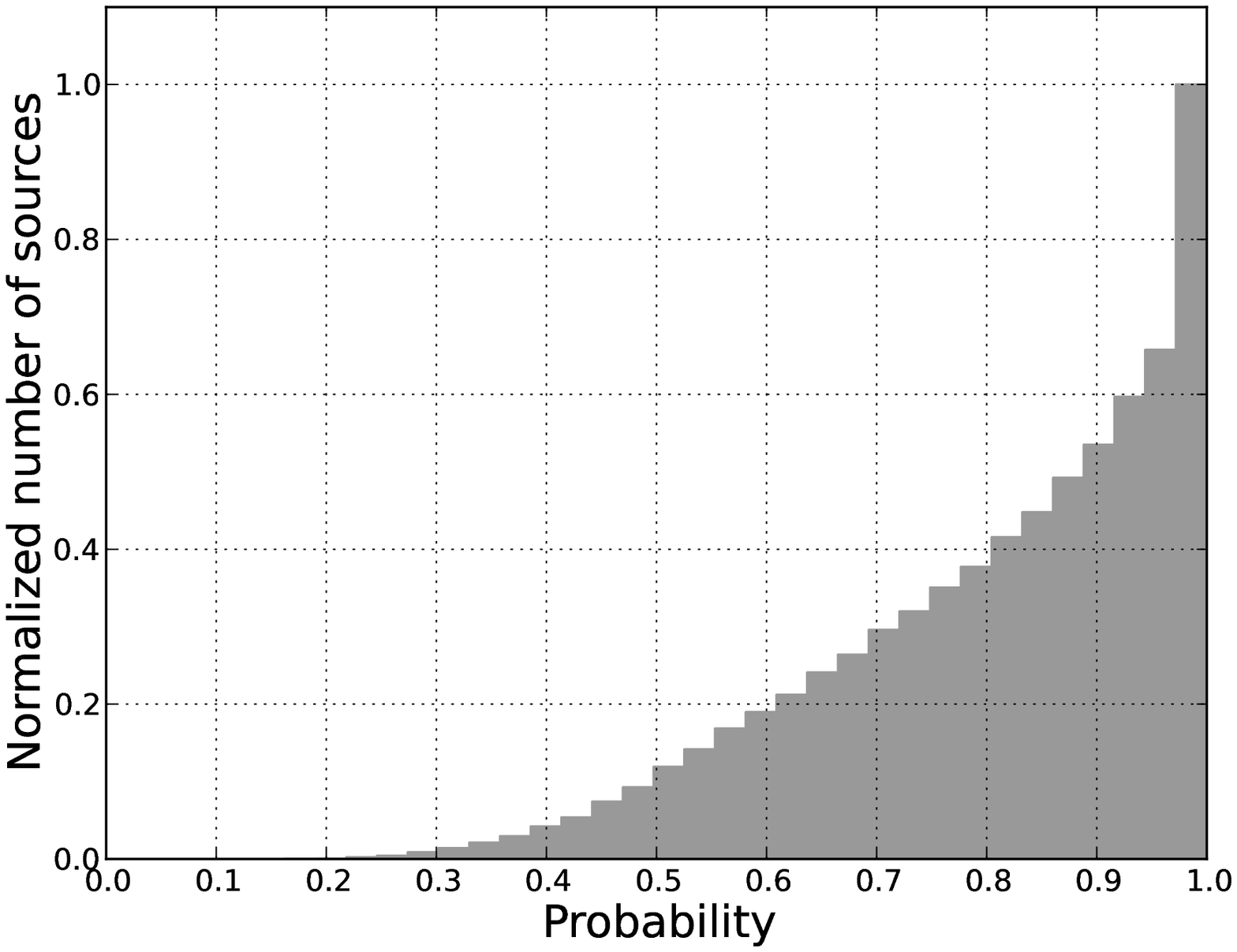}
\end{center}
    \caption{Cumulative histogram of probabilities of related class for $S_2$.
    There are more high-probability
    than low-probability light curves. 
    About half of the light curves have probabilities higher than 0.9.
    }
    \label{fiig:test_set_prob}
\end{figure}

In Fig. \ref{fiig:high_low_probability_light_curve}
we show two example EB light curves.
The probability of being in the EB class for the light curve in the left panel
is 1.0, while the probability for the light curve in the right panel is 0.38.
As the figure clearly shows, the light curve with a high probability
has a much stronger signal in the raw- and phased-folded light curve
than the one on the right. The estimated periods for EBs
are probably half of the true periods, particularly when 
there is little difference between amplitudes of primary and secondary eclipses.
However, it would not significantly affect the classification performance
because the model is trained not only on period, but other variability features as well.
In Fig. \ref{fiig:test_set_prob}, we
show the distribution of probability of related class for $S_2$ 
(probability, for short).
About half of the light curves' probabilities are higher than 0.9.
Conversely, there are fewer
samples with  low probabilities.
This is because we visually removed
light curves with no or feeble variability when constructing the training set.

The final model is trained using the 22 variability features,
the entire training set consists of 28\,392 light curves (i.e., $S_1 + S_2)$
and all the variable subclasses, where $t$ = 160 and $m$ = 13.
This model was used to select and classify 
periodic variable candidates from the EROS-2 database.

\subsubsection{Evaluation of the trained model}
\label{sec:sanity}

As mentioned in Section \ref{sec:training_set},
we removed 65\,229 (i.e., 97\,912 - 32\,683; see Table \ref{tab:training_set}) 
sources from the training set by visual examination.
Thus it is possible that we accidentally excluded true variables with low-level variability.
In this section, we assess the classification quality of
the trained model for weak variability using these visually removed sources.

\begin{figure}
\begin{center}
       \includegraphics[width=0.5\textwidth]{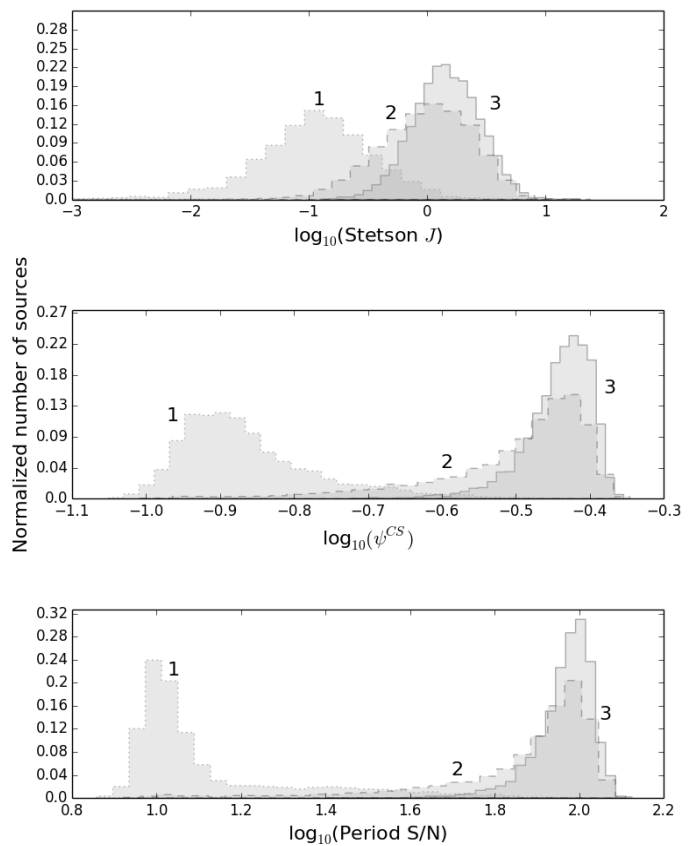}
\end{center}
    \caption{Histograms of three variability features, Stetson $J$, $\psi^{CS}$, and period S/N
    (from top to bottom).
    Each panel shows three histograms,
    each of which is a histogram of 
    1) non-variables (dotted line) in the training set,
    2) the visually removed RRLs (dashed line), and
    3) RRLs (solid line) in the training set.
    See text for details.
    }
    \label{fig:visually_removed_compare_histogram_RRL}
\end{figure}

\begin{figure}
\begin{center}
       \includegraphics[width=0.5\textwidth]{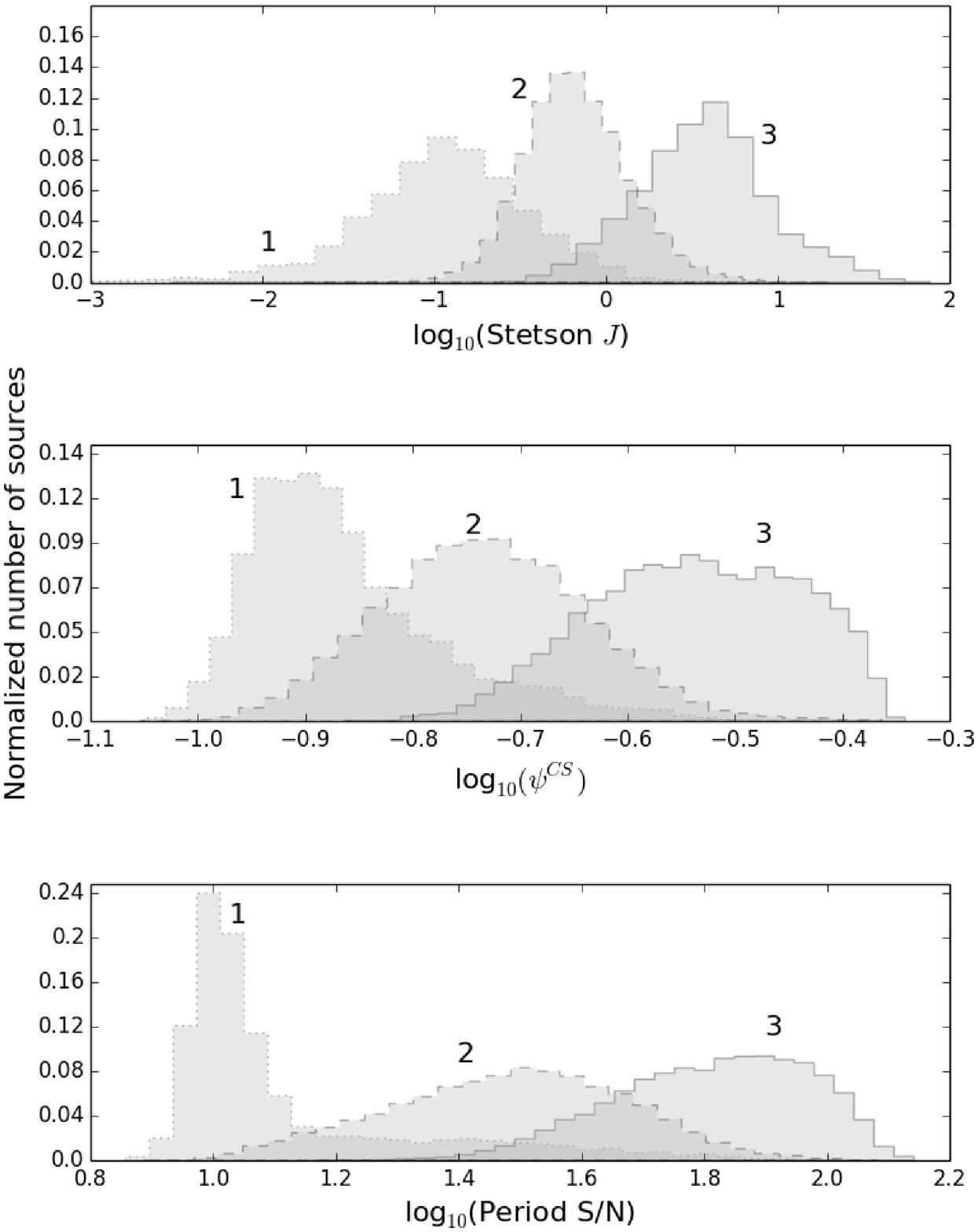}
\end{center}
    \caption{Histograms of three variability features, Stetson $J$, $\psi^{CS}$, and period S/N
    (from top to bottom).
    Each panel shows three histograms,
    each of which is a histogram of 
    1) non-variables (dotted line) in the training set,
    2) the visually removed LPVs (dashed line), and
    3) LPVs (solid line) in the training set.
    See text for details.
    }
    \label{fig:visually_removed_compare_histogram_LPV}
\end{figure}

We first refined these 65\,229 light curves
as explained in Section \ref{sec:reduction}, which yielded  38\,201 sources.
Most of the light curves were removed by the period criterion.
Fig. \ref{fig:visually_removed_compare_histogram_RRL} shows
histograms of variability features of 
1) non-variables in the training set, 
2) RRLs in the visually removed 38\,201 sources, and
3) RRLs in the training set.
The set of RRLs that were removed is
almost entirely overlapped by the RRL training set, implying that this
visual removal has not significantly biased the training set.
Fig. \ref{fig:visually_removed_compare_histogram_LPV} shows histograms
of the variability features of LPVs. 
LPVs that were
visually removed lie between the non-variables and the final LPV
training set. Objects in the left part of the histogram for the
removed LPVs have parameters similar to those of the non-variables in
the training set,  which means that they are probably also non-variables.  
The right half is similar to the LPVs in the training set, which indicates that they are
probably weakly variable LPVs mistakenly excluded by the visual
inspection (an inevitable consequence of trying to build a very clean
training set). 
Nevertheless, as the figure shows,  the training-set LPVs cover the right
half of the histogram well. We have
confirmed that other types of variables show a similar distribution to
either Fig. \ref{fig:visually_removed_compare_histogram_RRL} or
\ref{fig:visually_removed_compare_histogram_LPV}.

\begin{figure}
\begin{center}
       \includegraphics[width=0.5\textwidth]{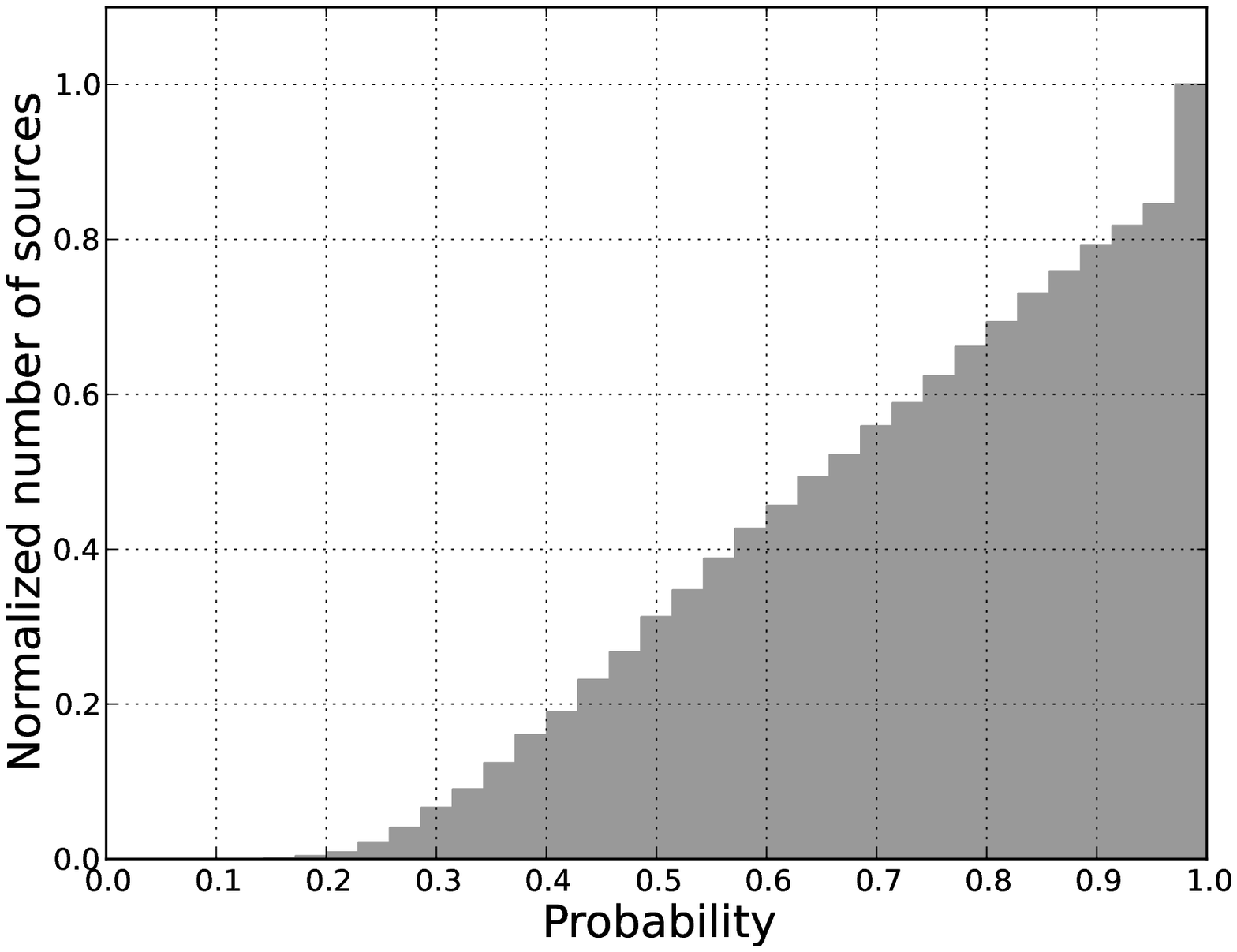}
\end{center}
    \caption{Cumulative histogram of probabilities of the visually removed sources
    that are identified as variable stars by the trained model.
    Compared with Fig. \ref{fiig:test_set_prob}, there are
    more  low-probability light curves.}
    \label{fiig:model_robust_prob}
\end{figure}

We then applied the trained model to these  38\,201 sources.
Among these sources, 12\,875 sources are classified as non-variables
and the remaining 25\,326 sources are classified as variables.
Fig. \ref{fiig:model_robust_prob} 
shows the probability distribution of the variable sources.
A relatively larger portion of
sources has low probabilities than in Fig. \ref{fiig:test_set_prob}.
This is not surprising  since these  sources
probably have weak variability and thus low probabilities.
Table \ref{tab:superclass_match_miss_match} shows
recall and precision for superclass classification.
The recall of CEPH is much worse than others.
Most of the CEPHs are classified as RRL c type stars
that generally show a similar variability pattern with CEPHs 
- slow increase and decrease of flux during its phase (i.e., sinusoidal).
We visually examined these misclassified CEPHs and
found that their light curves are not clearly distinguishable from RRLs.
Nevertheless, the high recall and precision of other periodic variable classes
indicates that the trained model is successful in classifying weak variability sources.
We also show recall and precision for the subclass classification of these variables
in Table \ref{tab:subclass_match_miss_match}.
The recall and precision of EBs and LPVs are poorer than others,
which might be associated with either
1) misclassification due to weak variability, or
2) incomplete OGLE subclass classification (e.g. see \citealt{Soszynski2009AcA, Graczyk2011AcA}).

\renewcommand{\arraystretch}{1.2}
\begin{table}
\small
\begin{center}
\caption{Superclass recall and precision for the removed sources\label{tab:superclass_match_miss_match}}
\begin{tabular}{ccc}
\hline\hline
Class & Recall & Precision \\
 &(\%) & (\%) \\
\hline
DSCT & 100.0 & 80.3 \\
CEPH & 20.7 & 37.5 \\
RRL  & 94.7 & 98.6 \\
EB & 96.4  & 93.0 \\
LPV & 99.7 & 99.9 \\
\hline
\end{tabular}
\end{center}
\end{table}

\renewcommand{\arraystretch}{1.2}
\begin{table}
\small
\begin{center}
\caption{Subclass recall and precision for the removed sources\label{tab:subclass_match_miss_match}}
\begin{tabular}{cccc}
\hline\hline
Superclass & Subclass & Recall & Precision \\
& & (\%) & (\%) \\
\hline
CEPH & & & \\
& 1O &  12.5  & 42.9 \\
& Other & 0.0 & 0.0 \\
RRL & & & \\
& ab & 93.3 & 99.4 \\
& c & 90.1 & 87.3 \\
& d & 70.6 & 86.0\\
& e & 80.2 & 74.1 \\
EB & & & \\
& EC & 40.3 & 26.5 \\
& ED & 85.6 & 75.4 \\
& ESD & 43.0 & 56.1\\
& ED+ESD & 7.3 & 10.2 \\
& Other & 0.0 & 0.0 \\
LPV & & & \\
& OSARG AGB C & 82.7 & 11.5 \\
& OSARG AGB O & 76.4 & 50.5 \\
& OSARG RGB C & 15.8 & 0.4 \\
& OSARG RGB O & 37.7 & 95.2 \\
& SRV AGB C & 38.3 & 38.6 \\
& SRV AGB O & 75.3 & 16.2 \\
\hline
\end{tabular}
\end{center}
Notes: We do not show some subclasses (e.g. CEPH F) 
whose recall or precision we were unable to calculate
because of the lack of the subclasses 
either in the OGLE classification
or in the trained model's classification. 
\end{table}

We visually examined the light curves 
with probabilities higher than 0.9 (see  Fig. \ref{fiig:model_robust_prob})
and found that the majority  of them show strong variability,
but were mistakenly excluded during the visual removal.
We trained another classification model after including
these high-probability sources into the training set to
check whether or not these samples help improve the model performance.
We found that the newly trained model showed almost identical performance to
the original model.

In this section, we assessed the classification quality of the trained model for weak variability. 
Although some true variables might be mistakenly removed during the visual inspection
because of the low level of variability,
most of the removed sources are relatively well covered
by the training set, as shown in Fig. \ref{fig:visually_removed_compare_histogram_RRL}
and \ref{fig:visually_removed_compare_histogram_LPV}.
Moreover, Tables \ref{tab:superclass_match_miss_match} 
and \ref{tab:subclass_match_miss_match} show that
the model has a fairly acceptable classification quality for such sources.
Nevertheless, note that the goal of this work is to select variables
from the 29 million EROS-2 sources where the majority
of the sources are expected to be non-variables.
Therefore minimizing possible false positives is the most critical task.
As mentioned in Section \ref{sec:training_set}, 
we observed a substantial decrease of classification performance without this visual removal.

\section{Variable candidates from the EROS-2 LMC database}
\label{sec:variable_candidate_selection}

We refined all light curves as depicted in Section \ref{sec:reduction} 
from the entire EROS-2 LMC database\footnote{About 50\%
of the EROS-2 light curves were removed by the criteria. 
Five sixths of them were removed by  the period criterion.},
extracted 22 variability features from them, and applied the trained model.
This yielded 150\,115 variable candidates.
From these light curves, we removed light curves with $B_E$
fainter than 20, which is around the limiting magnitude.
27\,224 light curves were removed by the magnitude cut.
We also visually examined some of the variable-candidate light curves
and found that some light curves with
a period S/N lower than 20 are likely false positives (i.e., non-variables).
For instance, the top-left panel of Fig. \ref{fig:features_plot} in the appendix
shows a scatter plot of period and period S/N
of the training-set light curves (symbols).
The panel also shows a contour line of $\sim$550k field sources from the EROS-2 LMC database.
The majority of the non-variable sources (gray circles) in the training set
and the field sources have a period S/N lower than 20.
Figs. \ref{fig:visually_removed_compare_histogram_RRL}
and \ref{fig:visually_removed_compare_histogram_LPV} show the same distribution.
Low-period S/N can be caused by large photometric uncertainties, 
relatively weak periodic signals, insufficient number of measurements, etc.
Although not every source with $B_E$ > 20 or period S/N $<$ 20 is a non-variable,
we removed all such sources to minimize the number of potential false positives.\footnote{
The training set light curves of variable sources show strong variability regardless
of their $B_E$ magnitudes or period S/N.}
5\,657 light curves were removed by the period S/N cut.
After this removal, the number of variable candidates is 117\,234,
which is about 0.44\% of the total sources in the EROS-2 LMC database.
In Table \ref{tab:variable_candidates_cut_20},
we show the number of variable candidates after these cuts.

\renewcommand{\arraystretch}{1.2}
\begin{table}
\small
\begin{center}
\caption{Number of variable candidates selected from the EROS-2 LMC fields \label{tab:variable_candidates_cut_20}}
\begin{tabular}{cccc}
\hline\hline
Superclass & Subclass & C1$^a$ & C2$^b$ \\
\hline
DSCT & & 2481 & 1906 \\
CEPH & &  \\
		& F &  1993 & 246 \\
		& 1O & 1527 & 318  \\
		& Other &  343 & 74 \\
RRL & & \\
		& ab  & 16407 & 3599  \\
		& c &  5367 & 1904 \\
		& d & 968 & 311  \\
		& e & 1636  & 793 \\
EB  & & \\
		& EC & 6120 & 5159  \\
		& ED & 13372 & 7216 \\
		& ESD & 13970 & 10627 \\
		& ED+ESD  & 674 & 422   \\
		& Other   & 11471 & 11138 \\
LPV  & & \\
		& Mira AGB C & 918  & 313 \\
		& Mira AGB O & 454 & 99  \\
		& OSARG AGB C   & 3538 & 883 \\
		& OSARG AGB O   & 15192 & 3992  \\
		& OSARG RGB C  & 1015    & 217 \\
		& OSARG RGB O   & 8863 & 3031  \\
		& SRV AGB C & 3884 & 1122 \\
		& SRV AGB O    & 6712  & 1737  \\
T2CEPH & & 329 & 178   \\
\hline
Total & & 117234 & 55285  \\
\end{tabular}
\end{center}
$^a$ After applying $B_E \le$ 20 and period S/N $\le$ 20 cuts.\\
$^b$ After additional removal of the known OGLE and MACHO variables.
\end{table}

\begin{figure}
\begin{center}
       \includegraphics[width=0.5\textwidth]{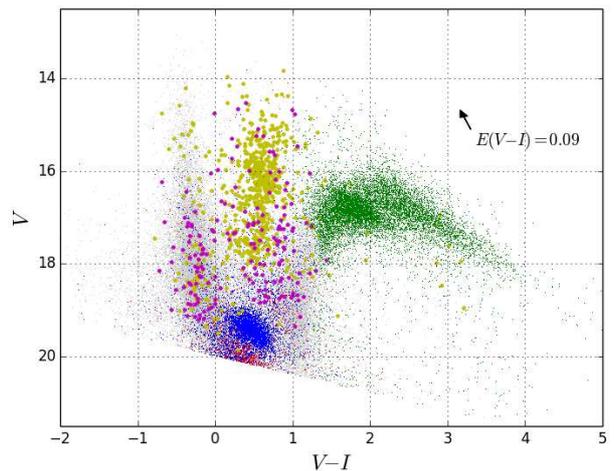}
\end{center}
    \caption{Color-magnitude diagram of the newly detected 55\,285 variable candidates.
    DSCTs: red dots, RRLs: blue dots, EBs: gray dots, LPVs: green dots,
    CEPHs: yellow circles and T2CEPHs: magenta circles.
    The magnitudes are not corrected for interstellar extinction.
    The arrow represents average interstellar extinction of 
    $E(V-I)$ = 0.09 for the LMC \citep{Haschke2011AJ}.
    The typical distance modulus to the LMC is 18.5 magnitude \citep{Tisserand2009AA}.
    A color version of this figure is available online.}
    \label{fig:new_variable_cmd}
\end{figure}

\begin{figure*}
\begin{center}
       \includegraphics[width=0.9\textwidth]{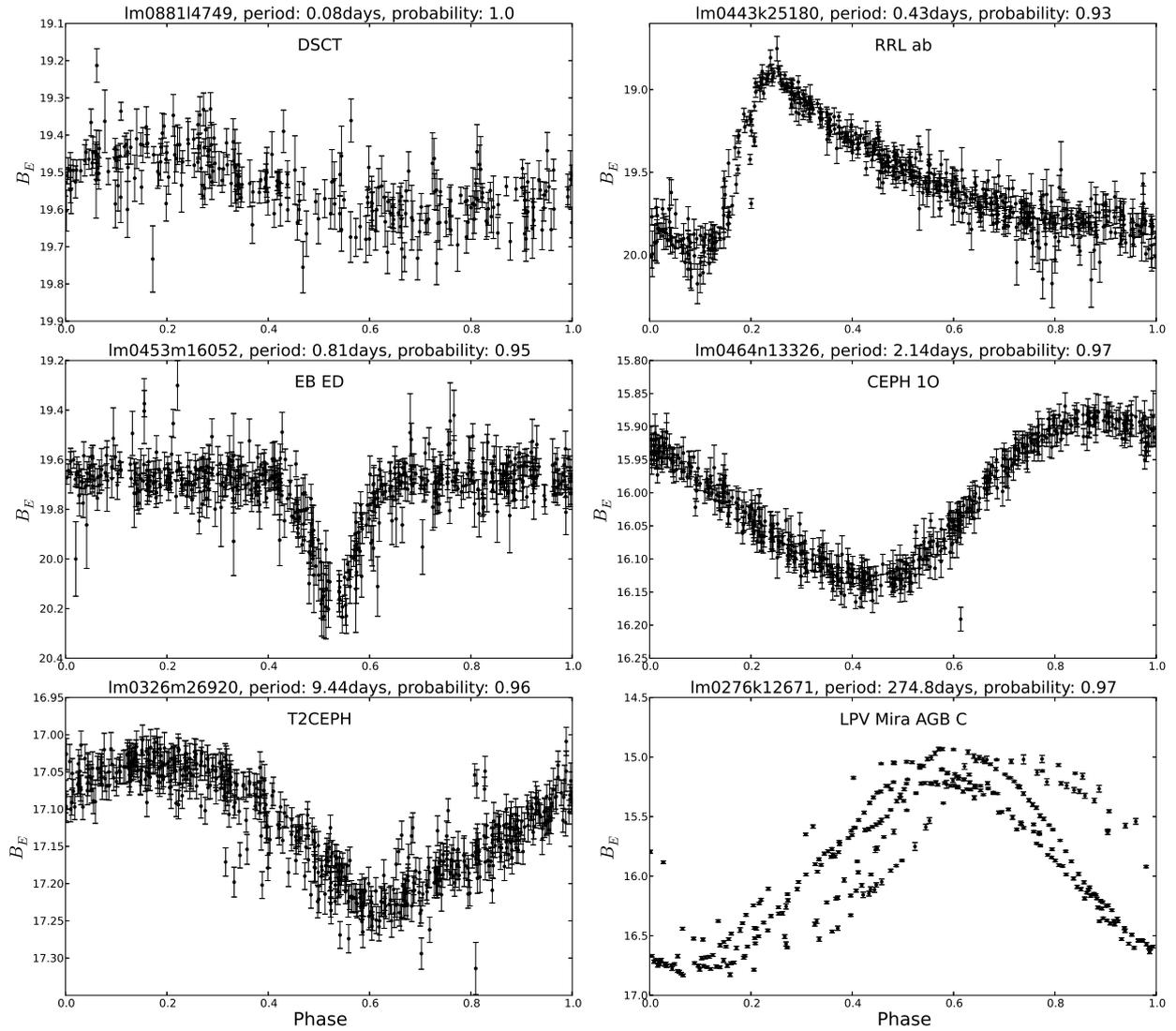}
\end{center}
    \caption{Examples of phase-folded light curves
    of six new variable candidates.}
    \label{fig:new_variable_candidates}
\end{figure*}

\begin{figure}
\begin{center}
       \includegraphics[width=0.5\textwidth]{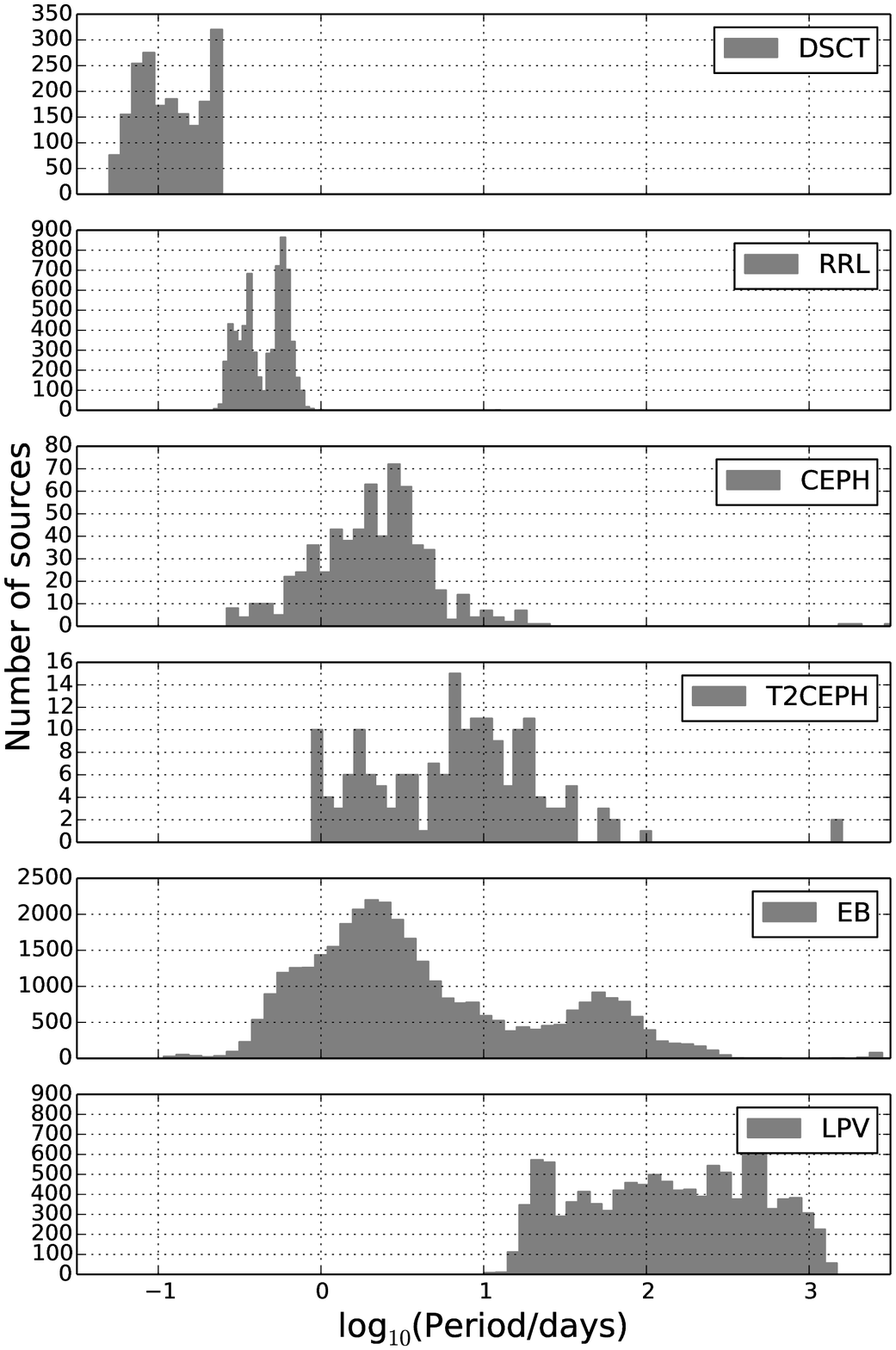}
\end{center}
    \caption{Period histograms of the new variable candidates.
    From top to bottom: DSCTs, RRLs, CEPHs, T2CEPHs, EBs, and LPVs.
    Each type of variable stars
    has a different period distribution.}
    \label{fig:new_variable_period_histogram}
\end{figure}

\begin{figure}
\begin{center}
       \includegraphics[width=0.5\textwidth]{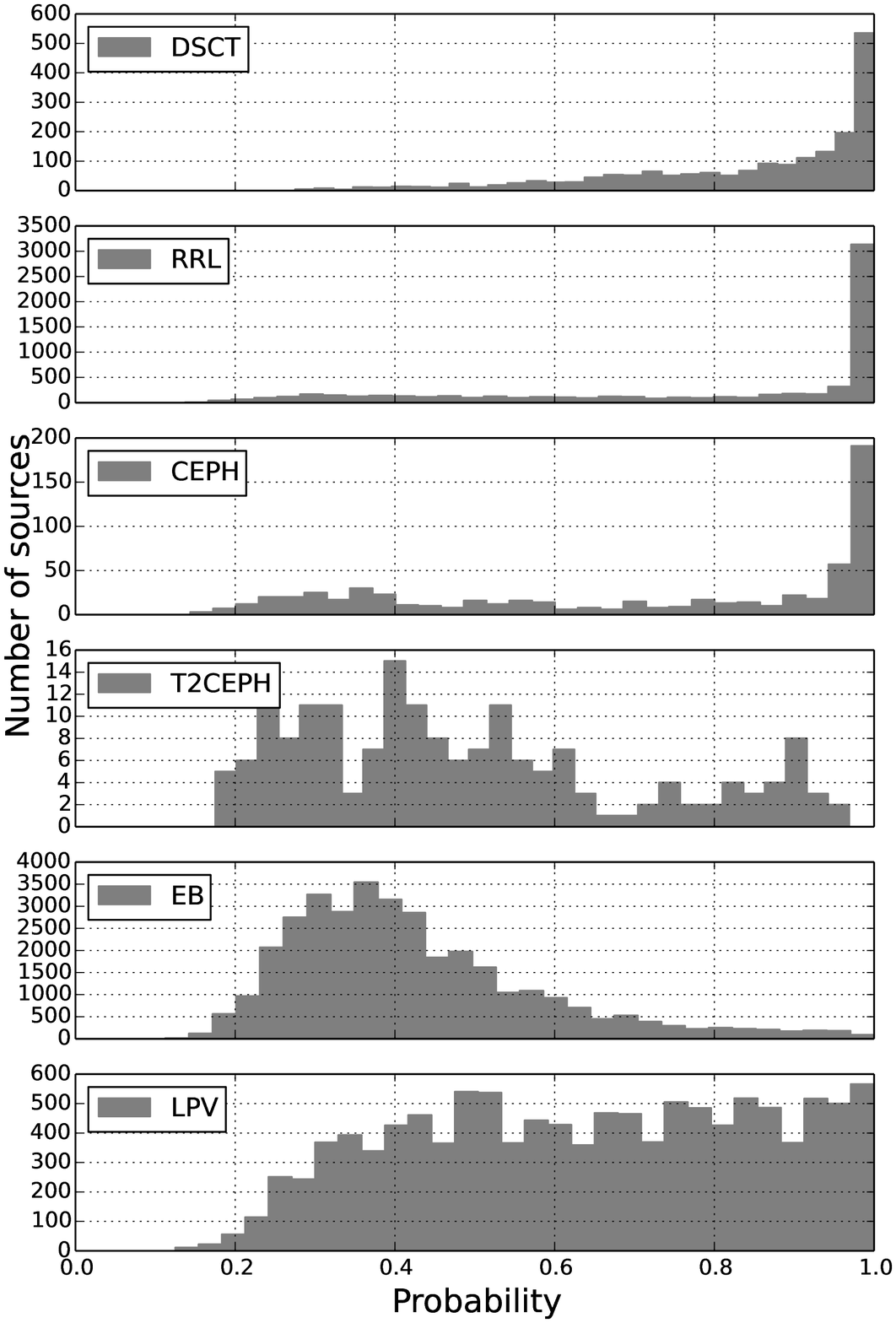}
\end{center}
    \caption{Probability histograms of the new variable candidates.
    Although most of DSCTs, RRLs, and CEPHs show a probability higher than 90\%,
    a relatively large potion of candidates of T2CEPHs, EBs, and LPVs show a probability lower than 0.9,
    which might be caused by misclassification within or between superclasses.}
    \label{fig:new_variable_prob_histogram}
\end{figure}

Among these 117\,234 variable candidates,
we found that 58\,069 variable candidates are  previously known OGLE variables.
We also crossmatched the 117\,234 EROS-2 variable candidates
with the known MACHO variables.
The list of known MACHO variables is taken
from \citet{Kim2011ApJ} and contains 19\,097 variables.
Among these 19\,097 MACHO variables, 
16\,543 are crossmatched with the EROS-2 variable candidates. 
We excluded EROS-2 DSCT and T2CEPH candidates
during the crossmatching because the list of the known MACHO variables
does not have these classes.
Table \ref{tab:variable_candidates_cut_20}
shows the number of variable candidates of each type
after removing the known OGLE and MACHO variables.
Fig. \ref{fig:new_variable_cmd}
is a color-magnitude diagram
of the new EROS-2 variable candidates.
We transformed EROS-2 bands $B_E$ and $R_E$
to the standard $V$ Johnson and $I$ Cousins broadband \citep{Tisserand2009AA}.
Fig. \ref{fig:new_variable_candidates}
shows examples of phase-folded light curves of six new variable candidates.
There is clear and strong variability
in the phase-folded light curves.
In Fig. \ref{fig:new_variable_period_histogram},
we show period histograms of each variable type.
It is well-known that different types of
periodic variable stars have different ranges of periods
such as DSCTs: 0.02 - 0.25 days \citep{Poleski2010AcA},
RRLs: 0.2 - 0.9 days \citep{Alcock1996AJ}
and CEPHs: days to months \citep{Soszynski2008AcA}.
We visually examined the two T2CEPHs with periods longer than 1\,000 days
and found that they might be misclassified LPVs. The probabilities
of these two T2CEPHs are relatively low, 18\% and 25\%.
The histogram of the EB periods in Fig. \ref{fig:new_variable_period_histogram} 
shows two populations, one with periods shorter than 20 days,
another with periods longer than 20 days.
We confirmed that the shorter period EBs are on the main sequence,
while the longer period EBs are on the red giant branch, which
is a known behavior \citep{Graczyk2011AcA}.
Fig. \ref{fig:new_variable_prob_histogram} shows the
probabilities of the candidates. 
Most of DSCTs, RRLs, and CEPHs have
probabilities higher than 90\%.
In contrast, many T2CEPHs, EBs, and LPVs
have probabilities lower than 50\%.
In particular, about two thirds of EBs have probabilities lower than 50\%,
which is probably caused by either misclassification within
EB subclasses or  misclassification between superclasses. 
Nevertheless, we visually examined these low-probability
candidates and confirmed that almost all of them show variability in their light curves.

\begin{figure}
\begin{center}
       \includegraphics[width=0.5\textwidth]{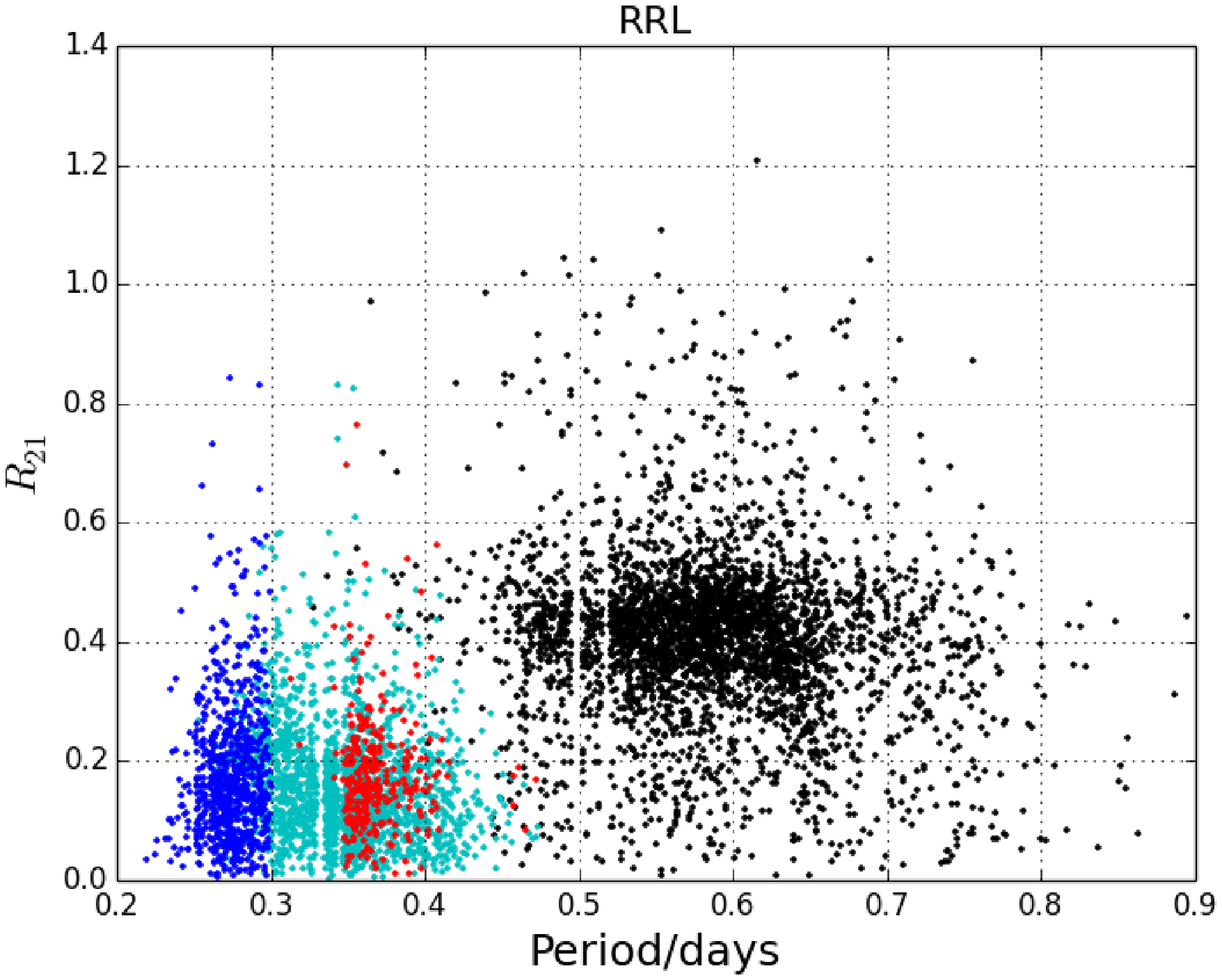}
\end{center}
    \caption{Period versus $R_{21}$ for the  new variable candidates of RRL type.
    RR ab: black circles, RR c: cyan circles,  RR d: red circles and RR e: blue circles.
    The vertical blanks are attributed to spurious periods.
    A color version of this figure is available online.}
    \label{fig:new_variable_RRL_feature}
\end{figure}

\begin{figure}
\begin{center}
       \includegraphics[width=0.5\textwidth]{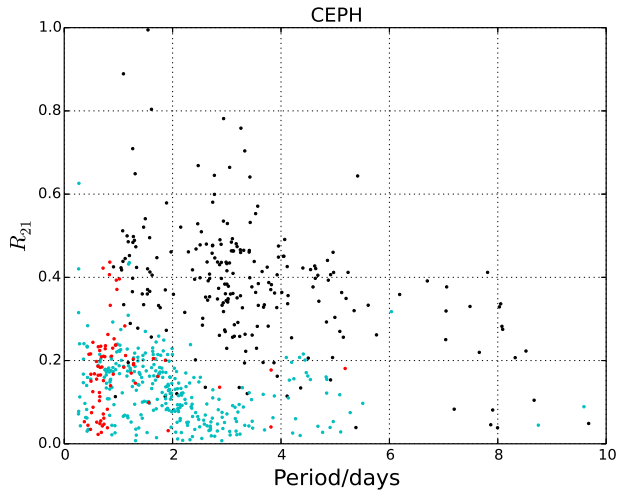}
\end{center}
    \caption{Period versus $R_{21}$ for the new variable candidates of CEPH type.
    F: black circles, 1O: cyan circles, and others: red circles.
    A color version of this figure is available online.}
    \label{fig:new_variable_CEPH_feature}
\end{figure}

\begin{figure}
\begin{center}
       \includegraphics[width=0.5\textwidth]{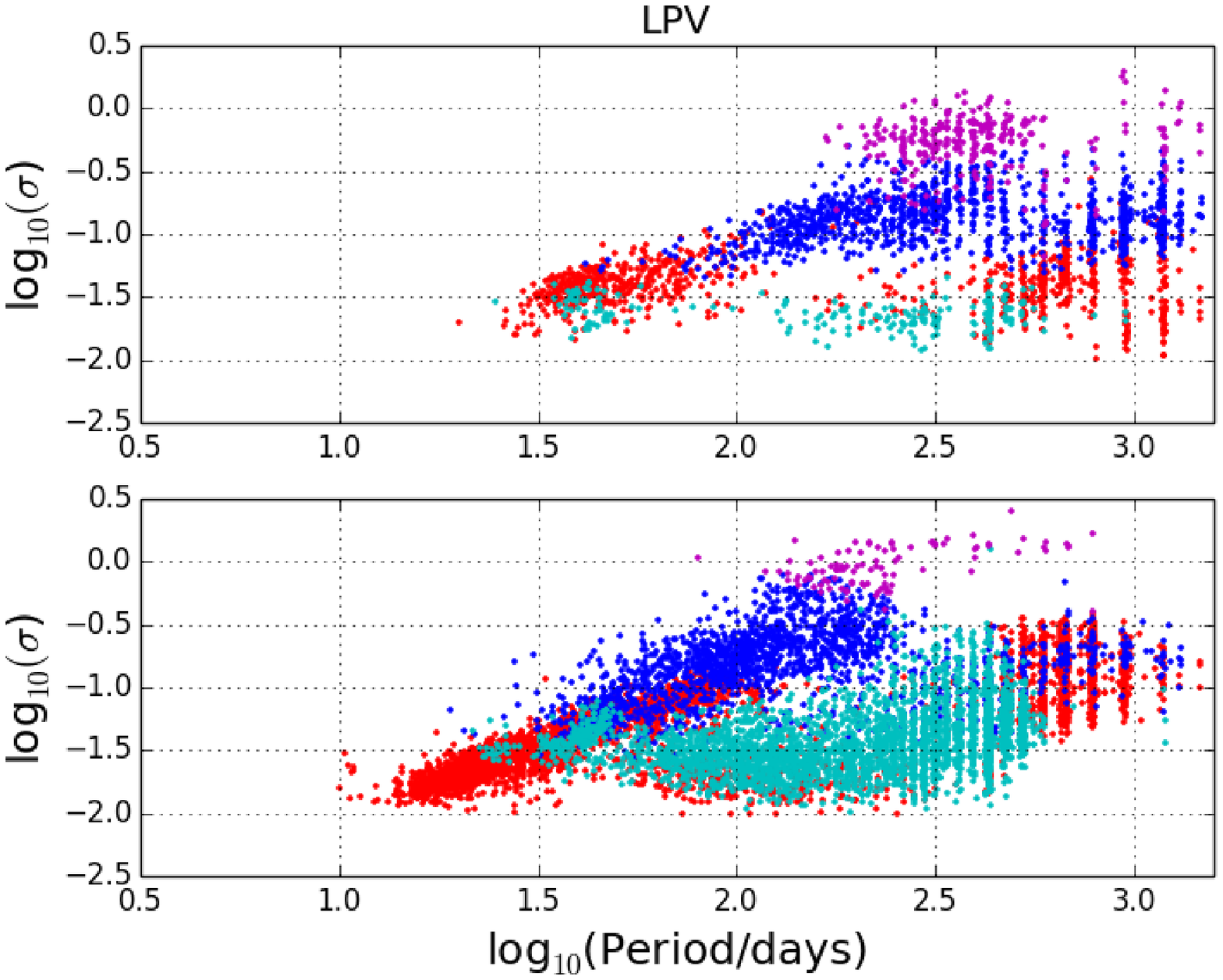}
\end{center}
    \caption{Period versus $\sigma$ for the new variable candidates of LPV type.
    OSARG AGBs: red circles, OSARG RGBs: blue circles, SRV AGBs: cyan circles and Mira AGBs: magenta circles.
    The carbon-rich LPVs (top panel) show relatively longer periods 
    than the oxygen-rich LPVs (bottom panel) \citep{Soszynski2009AcA}.
    Mira LPVs have larger amplitude than other LPVs,
    which is a known property of Mira variables \citep{Soszynski2009AcA, Soszynski2013AcA}.
    The vertical lines at longer periods are
    caused by a sparse sampling of the periodogram derived by the Lomb-Scargle method.
    A color version of this figure is available online.}
    \label{fig:new_variable_LPV_feature}
\end{figure}

Figs. \ref{fig:new_variable_RRL_feature} to
 \ref{fig:new_variable_LPV_feature}
show the relationships between features of 
each subclass of RRLs, CEPHs, and LPVs
in the newly detected 55\,285 variable candidates.
We do not show EBs since they are mixed
in most of the 2D-plane of variability features.
This is not surprising since the subclass classification performance 
of EBs is relatively poor (see Table \ref{tab:LibraryConfMatrix}).
Fig. \ref{fig:new_variable_RRL_feature} shows
period and amplitude ratio, $R_{21}$, of the RRL variable candidates;
see the appendix for details of $R_{21}$.
Each subclass is shown in a different color
and is distinguishable from each other, which is a
known property of RRL variables \citep{Soszynski2009AcA2}.
Fig. \ref{fig:new_variable_CEPH_feature}
shows the period and $R_{21}$ of CEPH variable candidates,
and each subclass is again relatively well distinguishable.
Fig. \ref{fig:new_variable_LPV_feature}
displays the relation of period versus $\sigma$ of the LPVs.

\section{Summary}
\label{sec:summary}

We presented the first result of the EPOCH project: 
the classification of periodic variable stars in the EROS-2 LMC light-curve database.
We first compiled the richest possible training set  based mainly on
the previously known OGLE variable stars.
We chose 22 variability features based on
the variable importance estimated by the random forest algorithm
and then calculated the features using the visually examined training set.
We then trained a random forest classification model using these variability features.
We applied the model to the 29 million EROS-2 LMC sources
and detected 117\,234 variable candidates.
The catalog of the variable candidates
containing EROS IDs, RA, Dec, colors (i.e., $B_E - R_E$), magnitudes (i.e., $B_E$),
periods, period S/N, probabilities and crossmatched OGLE/MACHO information
is available  at \href{http://stardb.yonsei.ac.kr}{http://stardb.yonsei.ac.kr}
and at the \href{http://vizier.u-strasbg.fr/viz-bin/VizieR}{CDS}.
Note that the catalog contains all 150\,115 variable candidates without removal of
the faint sources or low-period S/N sources mentioned in Section \ref{sec:variable_candidate_selection}.

The classification quality of any
supervised machine-learning methods
depends on the richness of the training set 
and  informativeness of the features on which a classification model is trained.
In this work, we used previously known OGLE variable sources to build a training set.
Thus a classification model would not be feasible 
for selecting and classifying variable types that 
do not exist in the OGLE variable source catalogs.
In future works, we will consider adding more variable sources of different types
to increase the completeness of the training set.
In addition, we visually removed 
sources without variability while building the training set,
which might result in an incomplete training set
because of unintended removal of weak-variability sources.
Although we showed that the trained model
was able to classify both strong and weak variability sources,
additional investigation on an enhanced training set
would be useful to increase classification quality.
We used 22 variability features
of the highest variable importance estimated with the random forest method.
We did not see any noticeable improvements by
using more or fewer features.
Nevertheless, it would be interesting
to perform a comprehensive feature selection 
based on a variety of methods (e.g. see \citealt{Guyon2003} and references therein)
to find irrelevant and/or highly correlated features
that could be removed without detriment to the classification quality.

In future works, we will apply a similar classification approach
to the one presented for the EROS-2 SMC, Galactic bulge, and spiral arm databases
to select and classify variable candidates.

\section*{Acknowledgements}

The EROS-2 project was funded by the CEA and the 
CNRS through the IN2P3 and INSU institutes. 
JBM  acknowledges financial support from "Programme National de Physique Stellaire" 
(PNPS) of CNRS/INSU, France.
Work at Yonsei was supported by the NRF grant 2012R1A1A2006924.
The analysis in this paper has been performed using the 
\href{http://hptc.fas.harvard.edu/}{Odyssey cluster} supported by the FAS Research 
Computing Group at \href{http://harvard.edu/}{Harvard}.

\bibliography{bibs}{}

\section*{Appendix}
\label{sec:appendix}

In this appendix, we explain each variability feature that we used to train the classification model.
$Q_{3-1|B-R}$, $\psi^{CS}$, $\psi^\eta$, $\eta_{B-R}$, and $\eta^e$  have been developed for this work.
Some other features were developed and/or used in other works
such as \citet{Shin2009MNRAS}, \citet{Kim2011ApJ}, \citet{Kim2012ApJ}, \citet{Pichara2012MNRAS} and \citet{Shin2012AJ}.

\begin{figure*}
\begin{center}
\begin{minipage}[c]{9cm}
        \includegraphics[width=1.0\textwidth]{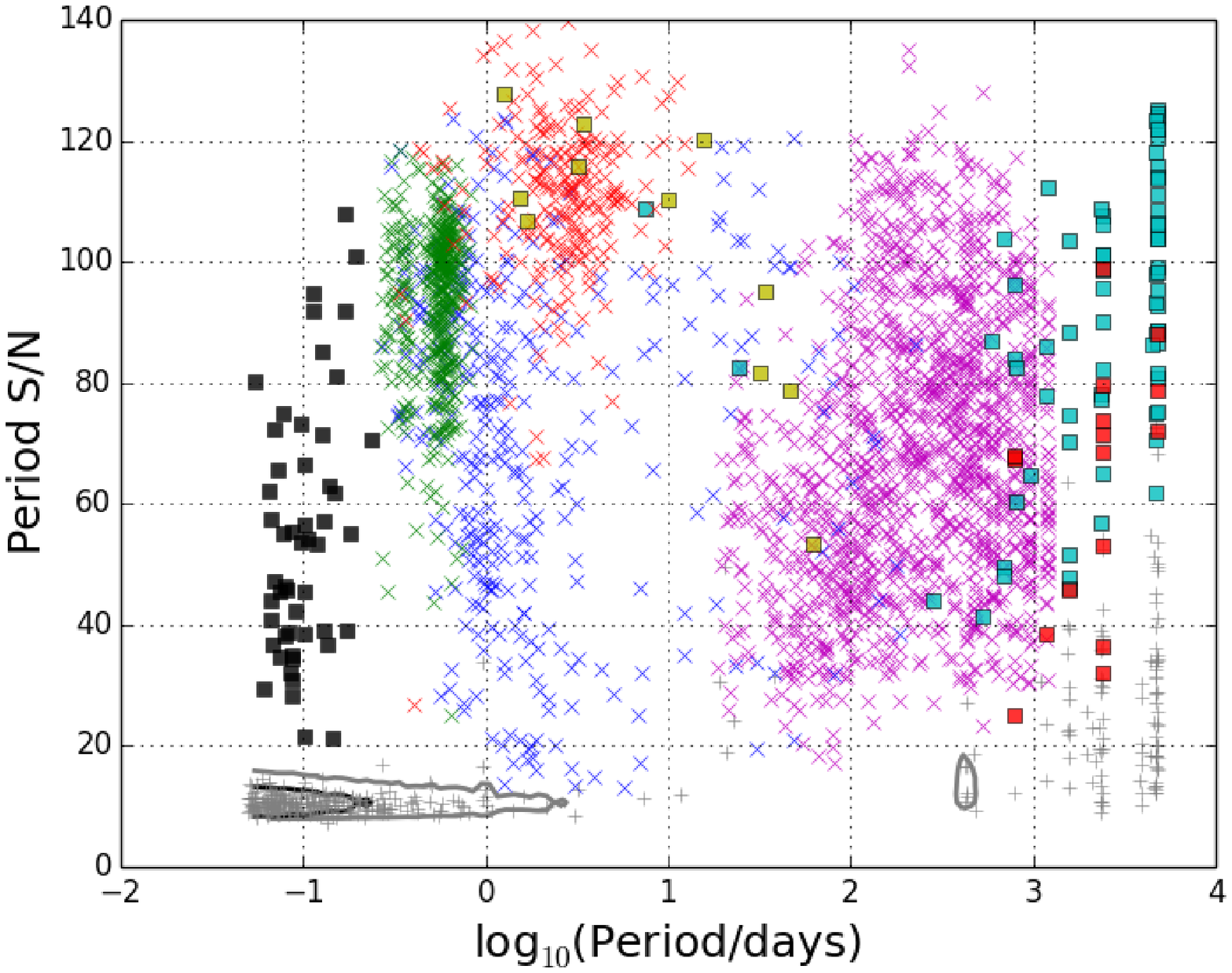}
\end{minipage}
\begin{minipage}[c]{9cm}
        \includegraphics[width=1.0\textwidth]{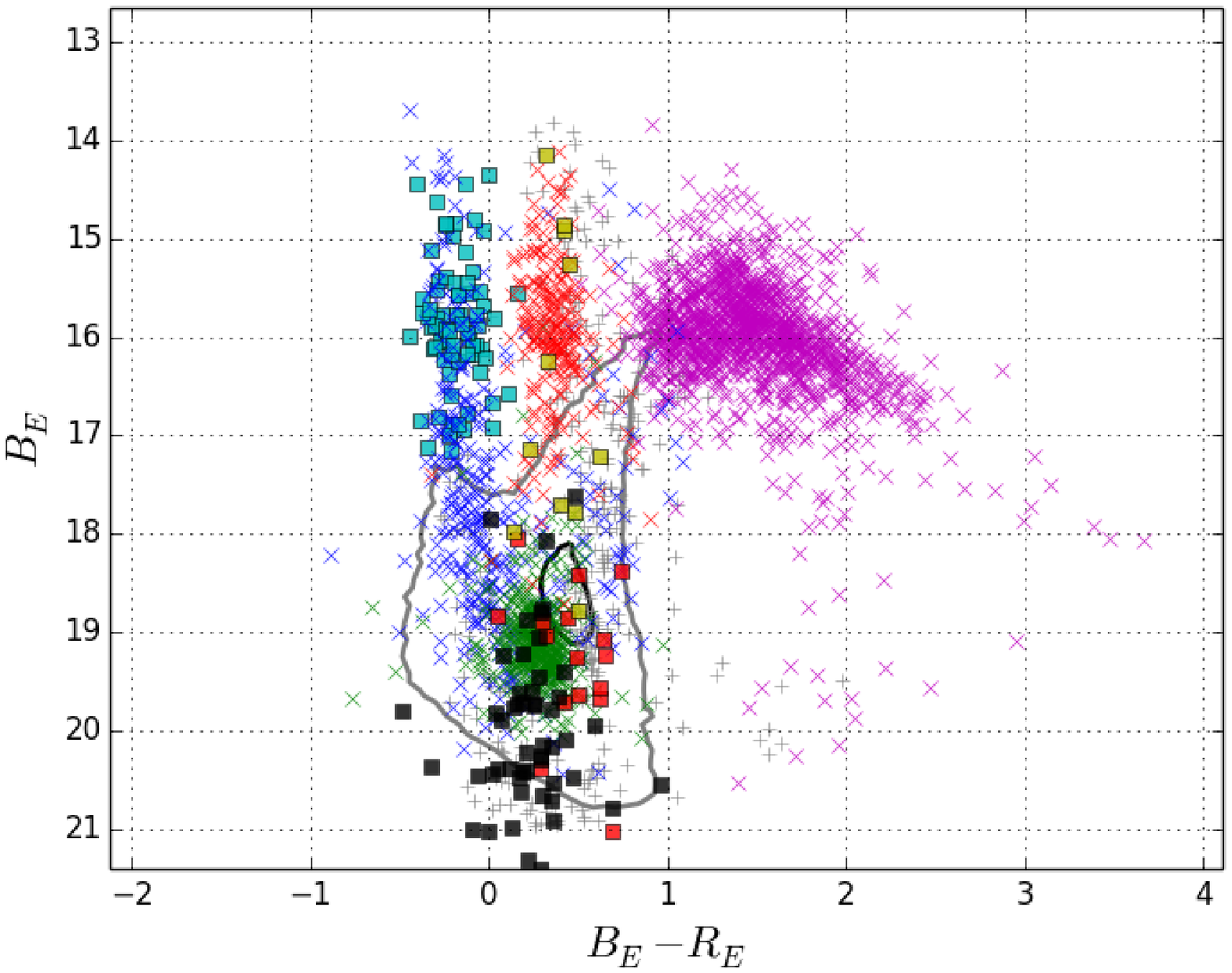}
\end{minipage}
\\
\begin{minipage}[c]{9cm}
        \includegraphics[width=1.0\textwidth]{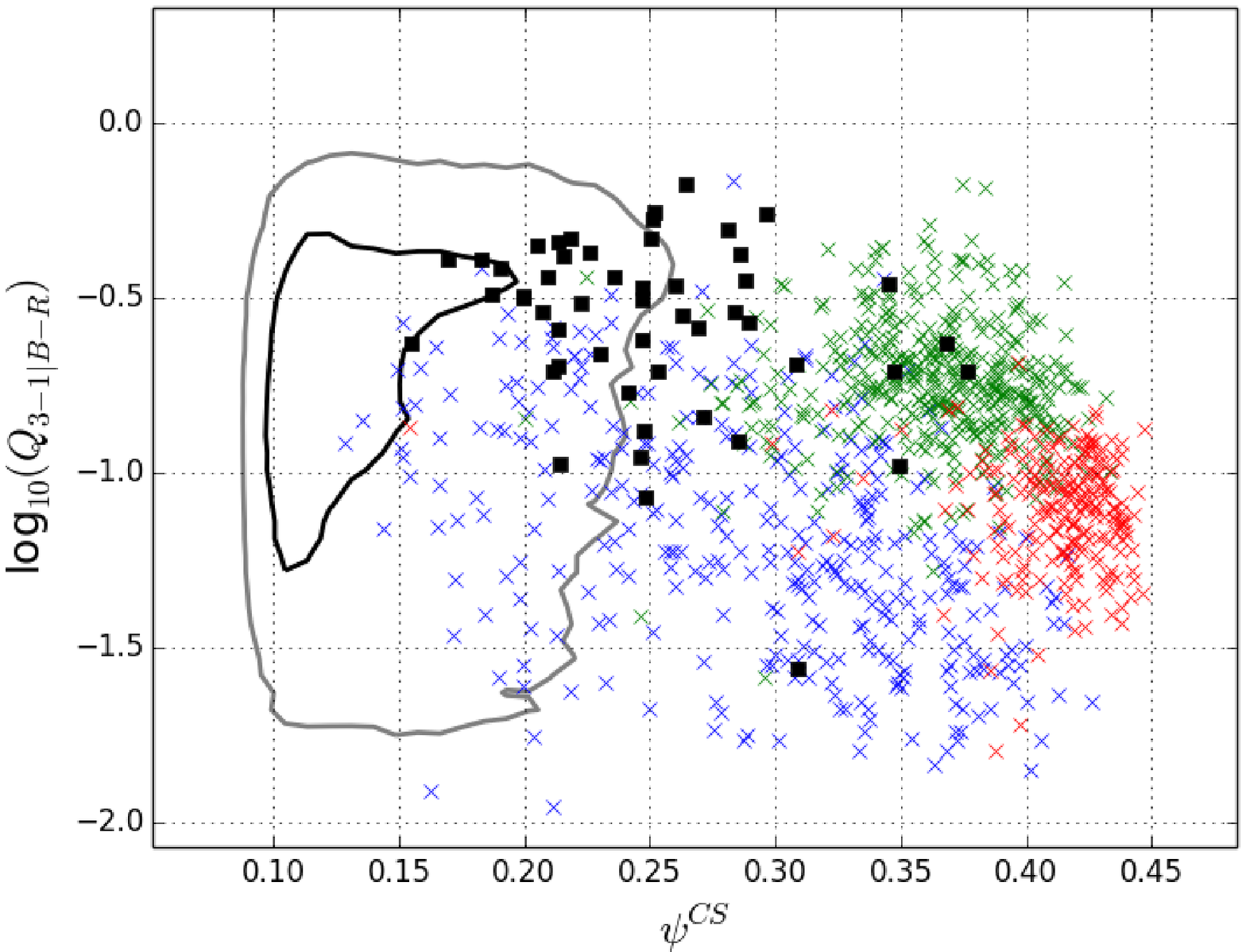}
\end{minipage}
\begin{minipage}[c]{9cm}
        \includegraphics[width=1.0\textwidth]{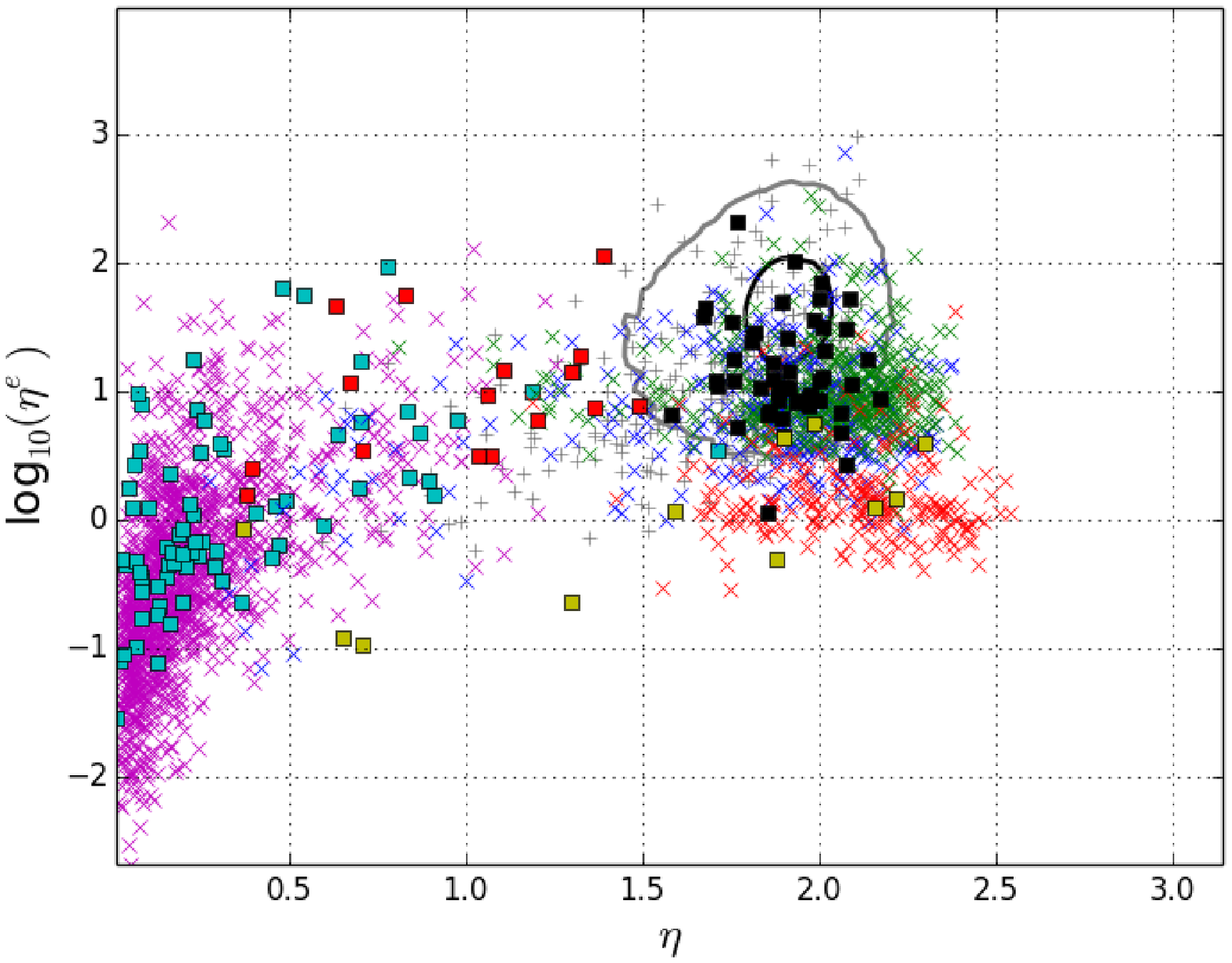}
\end{minipage}
\end{center}
    \caption
           {2D scatter plots of variability features of the training set explained in Section \ref{sec:training_set}.
	   DSCTs: black squares, RRLs: green x's,
           CEPHs: red x's, EBs: blue x's, 
           T2CEPHs: yellow squares,
           LPVs: magenta x's, QSOs: red squares, BVs: cyan squares, and NonVars: grey crosses.
           From left to right, clockwise: period versus period S/N,
           $B_E - R_E$ (i.e., color) versus $B_E$  band magnitude,
           $\eta$ versus $\eta^e$, and $\psi^{CS}$ versus $Q_{3-1|B-R}$.
           The contour line shows spatial distribution of about 550k field sources.
           To generate the contour line, we built a 2D histogram of 
           the field sources and then used the counts in each 2D-bin.
           The figure shows the two contour levels of 100 (thick gray line) and 1000 (thick black line).
           The majority of these field sources are probably non-variables.
           Different variable classes are separately grouped in the 2D space of variability features.
           We plot only one out of ten samples from the training set for better legibility of the figure.
           See the text for details.
           A color version of this figure is available online.}
    \label{fig:features_plot}
\end{figure*}

\begin{figure*}
\begin{center}
        \includegraphics[width=1.0\textwidth]{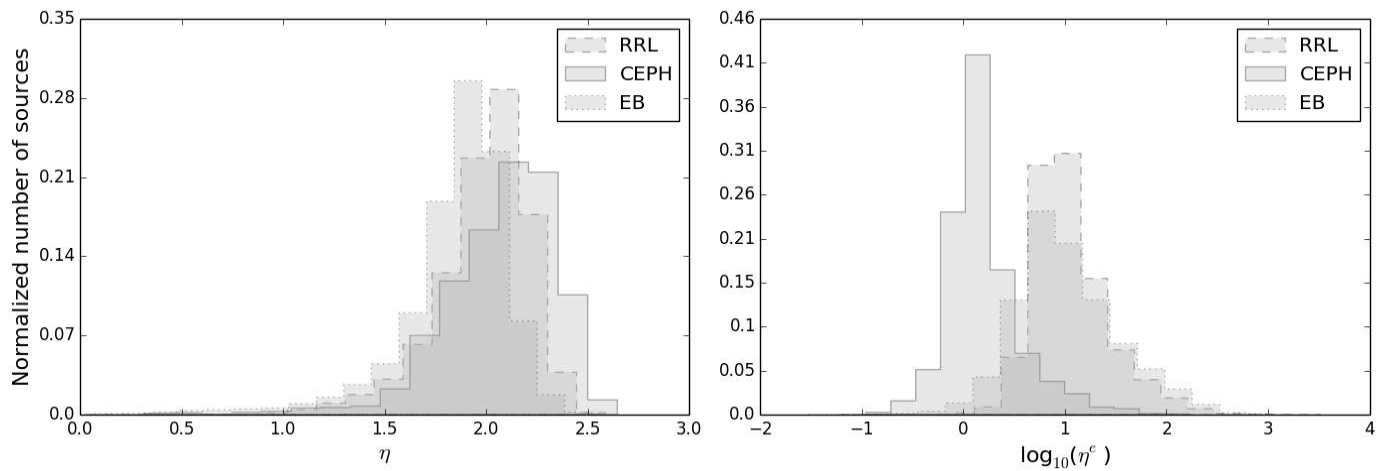}
\end{center}
    \caption
           {Histogram of $\eta$ (left panel) and
           $\eta^e$ (right panel) of RRLs (dashed), CEPHs (solid), and EBs (dotted).
           Using $\eta^e$, CEPHs are distinguishable from the other two classes.}
    \label{fig:eta_epoch_eta}
\end{figure*}

\begin{itemize}
 
\item Period and period S/N

In the top-left panel of Fig. \ref{fig:features_plot},
we show the period versus period S/N.
Most of the variable types are grouped in different regions.
For instance, LPVs (magenta x) have longer
periods than others, such as RRLs, EBs, CEPHs, and DSCTs.
The period was derived using the Lomb-Scargle algorithm \citep{Lomb1976ApSS}
and is the highest peak in the periodogram.

We define the signal-to-noise ratio (S/N) of the highest peak as
\begin{equation}
{\text{S/N}} = \frac{{\text{max}}(p_{LS}) - \bar{p}_{LS}} {\sigma_{p_{LS}}},
\end{equation}
where $p_{LS}$ is power of all frequencies in the periodogram,
max is the maximum value, $\bar{p}_{LS}$ is a mean power,
and  $\sigma_{p_{LS}}$ is a standard deviation.
Although the S/N of a peak \citep{Scargle1982ApJ} is generally 
defined as
\begin{equation}
{\text{S/N}} = \frac{{\text{max}}(p_{LS})} {\bar{p}_{LS}},
\end{equation}
{\noindent}we used the alternative definition since
we empirically found  that $\sigma_{p_{LS}}$ of
peridograms of EROS-2 light curves are relatively large,
which could be attributed to noise in the light curves such as systematic trends and
high photometric uncertainty.
The top-left panel of Fig. \ref{fig:features_plot} shows that
the majority of non-variables (gray circles) 
and the field sources (contour line) have a lower period S/N than other variables.

\item $\eta$

$\eta$ measures the degree of 
trends (e.g. monotonic increase or decrease of flux 
in a long-term baseline) \citep{Neumann1941}.
$\eta$ is defined as

\begin{equation}
\eta = \frac{1}{(N-1) \,\, \sigma^{2}} \sum_{i=1}^{N-1}(m_{i+1} - m_{i})^{2},
\end{equation}

{\noindent}where $N$ is the total number of measurements of a light curve,
$\sigma$ is the standard deviation, $m$ is the magnitude, and
$i$ is the index of time.
$\eta$ is known to be useful to separate
light curves with long-term variation,
as shown in the bottom-right panel of Fig. \ref{fig:features_plot}.

\item Color and magnitude

Color and magnitude
are powerful indices for variable separation because 
 different variable types are generally placed 
at different regions in a color-magnitude diagram (CMD).
The CMD of the training set is shown in the
top-right panel of Fig. \ref{fig:features_plot}.
Although there is some overlap between classes,
some classes are clearly distinguishable in the CMD, for example, LPVs.

\item $Q_{3-1}$

$Q_{3-1}$ is the difference between the third quartile, $Q_3$, and the first quartile, $Q_1$, of a raw light curve.
$Q_1$ is a split between the lowest 25\% and the highest 75\% of data.
$Q_3$ is a split between the lowest 75\% and the highest 25\% of data.

\item $Cusum$

$Cusum$ is the range of a cumulative sum (e.g. see \citealt{Ellaway1978}) of a light curve and is defined as

\begin{eqnarray}
\begin{array}{l}
Cusum = {\rm{max}}(S) - {\rm{min}}(S)
\\ \\
S_{l} = \displaystyle \frac{1}{N \,\, \sigma} \sum_{i=1}^{l}(m_{i} - \bar{m}),
\end{array}
\end{eqnarray}

{\noindent}where max (min) is the maximum (minimum) value of $S$ and $l=1, 2, \ldots, N$.
$\bar{m}$ is a mean magnitude.
$Cusum$ is generally high for 
light curves with long-term variability such as  LPVs, BVs, and QSOs,
while it is relatively low for light curves of short-periodic variables and non-variables \citep{Kim2011ApJ}.

\item $Q_{3-1|B-R}$ and  $\psi^{CS}$

$Q_{3-1|B-R}$ and  $\psi^{CS}$ are
not the features derived from a raw light curve
but the features derived from a $B_E - R_E$ light curve, $Q_{3-1|B-R}$, 
and from a phase-folded light curve, $\psi^{CS}$.
We generated a $B_E - R_E$ light curve using only the simultaneous measurements 
from $B_E$ and $R_E$.
If a variable star's flux variability is achromatic, 
the generated $B_E - R_E$ light curve follows the normal  distribution (i.e., Gaussian noise).
Otherwise it would show a certain variability in the $B_E - R_E$ light curve.

$\psi^{CS}$ is the range of the cumulative sum ($Cusum$) of
a phase-folded light curve.
A phase-folded light curve is generated using the period estimated from the Lomb-Scargle method.
For relatively short-period variables such as DSCTs, RRLs, CEPHs, and EBs,
a phase-folded light curve would have a different
shape from a raw light curve
and thus embrace different variability characteristics.
In the bottom-left panel of Fig. \ref{fig:features_plot}, we show the scatter plot
of  $Q_{3-1|B-R}$ versus $\psi^{CS}$ of
short-period variables including DSCTs, RRLs, CEPHs, and EBs.
The short-period variables are relatively well distinguishable.
We do not show other types of variables because
they are spread across the plane.

\item $\psi^\eta$

$\psi^\eta$ is the $\eta$ index 
calculated from the phase-folded light curve.
\citet{Dubath2011MNRAS} used a conceptually
identical feature with $\psi^\eta$, which they termed {\em{P2p scatter}}.

\item $\eta_{B-R}$

This is an $\eta$ index calculated from the  $B_E - R_E$ light curve.

\item $\eta^e$

Although $\eta$ is a powerful index for quantifying
variability characteristics of a light curve,
it does not take into account unequal sampling. 
Thus we define $\eta^e$ as

\begin{eqnarray}
\begin{array}{l}
\eta^e = \displaystyle \bar{w} \,\, (t_{N-1}-t_1)^2 \,\, \frac{\sum_{i=1}^{N-1} w_i \,\, (m_{i+1} - m_{i})^2} {\sigma^2 \,\, \Sigma^{N-1}_{i=1} w_i} 
\\ \\
w_i = \displaystyle \frac{1} {(t_{i+1} - t_i)^2},
\end{array}
\end{eqnarray}

{\noindent}where $t$ is the measurement epoch, $t_{i+1} > t_i$,
and $\bar{w}$ is the mean value of $w_i$.
The right panel in Fig. \ref{fig:eta_epoch_eta}
shows the $\eta^e$ histogram of RRLs, CEPHs, and EBs.
Although the separation between the classes is not perfect,
CEPHs are clearly distinguishable from others.
We designed $\eta^e$ considering
different sampling rates and baselines between light curves,
which is useful in some cases such as shown in 
Fig. \ref{fig:eta_epoch_eta} and also in the bottom-right panel of Fig. \ref{fig:features_plot}.
Additional investigation, however, would be necessary
to examine detailed characteristics of $\eta^e$.

\item $\sigma$, $\gamma_1$ and $\gamma_2$

$\sigma$ is a standard deviation, $\gamma_1$ is a skewness
and $\gamma_2$ is a kurtosis of a  light curve.
$\gamma_1$ and $\gamma_2$ are defined as

\begin{eqnarray}
\begin{array}{l}
\displaystyle
\gamma_1 = \frac{N} { (N -1) (N - 2)} \sum_{i=1}^N (\frac{m_i - \bar{m}} {\sigma})^3
\\ \\
\displaystyle
\gamma_2 =   \frac{N (N+1)} { (N -1)(N - 2)(N-3)} \sum_{i=1}^N (\frac{m_i - \bar{m}} {\sigma})^4  
\\
\displaystyle
\,\,\,\,\,\,\,\, \,\,\,\,\,- \frac{3(N-1)^2}{(N-2)(N-3)},
\end{array}
\end{eqnarray}

\item $J$ and $K$

$J$ is a Stetson $J$ index \citep{Stetson1996PASP}, which is
 calculated based on simultaneous measurements 
 from light curves in two bands, $p$ and $q$, of the same star. It is defined as

\begin{eqnarray}
\begin{array}{l}
\displaystyle
J = \frac{1}{N}\sum_{i=1}^{N}{\rm{sign}}(P_{i})\sqrt{|P_{i}|}
\\ \\
\displaystyle
P_{i} = \delta_{p}(i)\, \delta_{q}(i)
\\ \\
\displaystyle
\delta_{p}(i) = \sqrt{\frac{N}{N-1}} \, \, \frac{m_{p,i} - \bar{m}}{\sigma_{p,i}},
\end{array}
\end{eqnarray}

{\noindent}where $N$ is the total number of measurements, $i$ is an index for each measurement,
and sign($P_{i}$) is the sign of $P_{i}$.
$\sigma_{p,i}$ is the standard error of $i_{th}$ measurement of band $p$.
In case of the EROS-2 database, 
$p$ and $q$ indicate $B_E$ and $R_E$ bands.

Stetson $K$ is calculated using a single band light curve as

\begin{eqnarray}
K = \frac{1}{\sqrt{N}} \frac{\sum_{i=1}^{N}|\delta(i)|} {\sqrt{\sum_{i=1}^{N}\delta(i)^{2}}},
\label{eq:StetsonK}
\end{eqnarray}

A pure sinusoid signal has $K = 0.900$  and Gaussian distribution has $K = 0.798$.

\item $K_{AC}$

$K_{AC}$ is Stetson $K$ index calculated over
the autocorrelation function of a light curve.
This feature is useful to separate
long- or non-periodic variables including QSOs from periodic variables 
(see \citealt{Kim2011ApJ} for details).

\item $R_{21}$, $R_{31}$, $\phi_{21}$, and $\phi_{31}$

These features are derived using the Fourier decomposition \citep{Petersen1986AA, Grison1994AA}.
$R_{k1}$ is an amplitude ratio and $\phi_{k1}$ is a phase difference,
which is defined as

\begin{equation}
R^2_{k1} = \frac{A^2_k + B^2_k} {A^2_1 + B^2_1}
\end{equation}
\begin{equation}
\phi_{k1} = \phi_k - k\phi_1, \text{ where } \phi_k = tan^{-1}(-\frac{B_k}{A_k}).
\end{equation}

To calculate $A_k$ and $B_k$, we adopted the Grison model-fitting procedure \citep{Grison1994AA},
which first constructs a zero-mean light curve
and then fits a Fourier series, defined as

\begin{equation}
\sum^M_{i=1} [A_i cos(i \frac{2\pi}{P}t) + B_i sin(i \frac{2\pi}{P}t)],
\end{equation}

{\noindent}where M is the number of series to fit,
$P$ is the derived period from the Lomb-Scargle method, and $t$ is the time epoch.
The features $R_{k1}$ and $\phi_{k1}$ can
distinguish different shapes of light curves 
and thus are useful to separate subclasses of 
short-period variables such as CEPHs \citep{Beaulieu1995AA}.

\end{itemize}

\end{document}